%% file: main_v3_revised.tex
\documentclass[11pt]{article}

\usepackage[utf8]{inputenc}
\usepackage[T1,T2A]{fontenc} 
\usepackage[a4paper,margin=1in]{geometry}
\usepackage{microtype}

\usepackage{amsmath, amssymb, amsfonts}
\usepackage{booktabs}
\usepackage{multirow}
\usepackage{array}
\usepackage{graphicx}
\usepackage{subcaption}
\usepackage{float}
\usepackage[section]{placeins} 
\usepackage{xcolor}
\usepackage{enumitem}
\usepackage{url}
\usepackage{tikz}
\usetikzlibrary{arrows.meta, positioning, calc}
\usepackage[hidelinks]{hyperref}
\usepackage[nameinlink,capitalise,noabbrev]{cleveref}

\usepackage{makecell}
\usepackage{adjustbox}
\usepackage{footnote}
\usepackage{fvextra} 
\makesavenoteenv{table} 

\newcolumntype{L}[1]{>{\raggedright\arraybackslash}p{#1}} 

\newcommand{\figbox}[1]{\includegraphics[width=\linewidth]{#1}}

\tikzset{
  fbox/.style={draw, rounded corners, minimum height=9mm, align=center, font=\small},
  fguard/.style={draw, fill=black!6, rounded corners, minimum height=9mm, align=center, font=\small},
}

\title{\textbf{HiveTraceGuard-Pro: A Compact Generative Guardrail for Prompt Injection, Jailbreaks, and Adversarial Obfuscation}}
\author{
\begin{tabular}{c@{\hspace{9mm}}c@{\hspace{9mm}}c}
\textbf{Nikita Oblakov} & \textbf{Sabrina Sadiekh} & \textbf{Evgeniy Kokuykin} \\
\textit{HiveTraceLab} & \textit{HiveTraceLab} & \textit{HiveTraceLab} \\
{\small\texttt{nikobl25@gmail.com}} & {\small\texttt{sadsobr7@gmail.com}} & {\small\texttt{evgeniy.kokuykin@raftds.com}}
\end{tabular}
}

\begin{document}

\maketitle

\begin{abstract}

Production LLMs must handle inputs that attempt to override system instructions, bypass safety policies or elicit harmful responses. A common mitigation is a separate guardrail model. Existing reports, however, provide little evidence on Russian prompt injection or Russian surface obfuscation. We present HiveTraceGuard-Pro, a 0.6B generative guardrail LoRA-tuned from Qwen3-0.6B. It is trained on Russian and English and uses one binary scoring rule (\texttt{safe}/\texttt{unsafe}) for the final target turn. Its training corpus pairs harmful examples, where a counterpart exists, with benign examples from the same domain and applies eight obfuscation transforms to both labels.

In one harness, we compare HiveTraceGuard-Pro with thirty-four other guards on nineteen benchmark groups, sixteen of which are public. Its aggregate key is 0.7432, behind 0.7641 and 0.7552 for the two higher-scoring guards. Over the sixteen public groups alone, its key is 0.7153 and four of the thirty-four other suite guards score higher. In a fifteen-model comparison, HiveTraceGuard-Pro has the highest clean Russian robustness combined-F1 (0.88) and Russian prompt-injection recall (0.999). Both results use Russian sets assembled by our team, and at least 27.1\% of the prompt-injection set overlaps the training corpus. Its 14.3\,ms median latency is the lowest among those fifteen models in that run. Across the suite, FPR is 0.268 and FNR is 0.156. All reported response results use a legacy standalone-reply serialization rather than the natural \texttt{assistant}-role path of the shipped chat template.

We release the merged weights on Hugging Face under Apache-2.0. The corpus, evaluation sets and evaluation code remain internal.

\end{abstract}

\section{Introduction}

Production language models interact with untrusted users. A guardrail can evaluate a user request before generation or an assistant response before delivery. Because the guard is separate from the assistant, its moderation behavior can be updated without retraining the assistant itself.

Published guardrails focus mainly on English in both training and evaluation. Russian-language attacks can combine instruction replacement with transliteration, keyboard-layout substitutions, symbol insertion or informal rewriting. Published evaluations provide limited evidence about how existing guards handle these combinations.

A guardrail serving Russian traffic must distinguish harmful intent from benign discussion of the same topic and preserve that distinction when the surface form changes. We also target deployments with tight latency and memory budgets. HiveTraceGuard-Pro therefore keeps the base model at 0.6B and uses a corpus weighted toward Russian, a structured harm taxonomy, paired benign and harmful examples, and obfuscation applied to both labels.

\paragraph{Contributions.}
\begin{itemize}\itemsep2pt
\item A compact (0.6B) generative guardrail trained on Russian and English that uses one model and one binary decision rule for requests and responses. The evaluation places particular emphasis on Russian robustness.
\item A corpus construction method that pairs harmful examples, where a counterpart exists, with benign examples from the same domain and applies obfuscation to both labels. This makes over-blocking measurable on topics that also contain harmful examples. We release the weights under Apache-2.0 and describe the method, but withhold the corpus (\cref{sec:availability}). The generative label-emission mechanism itself is established practice~\cite{qwenguard,llamaguard,shieldgemma}.
\item A confirmation set reserved before development and evaluated once after the model was frozen (\cref{sec:confirm}), together with a training-overlap audit of the scored benchmarks and overlap-adjusted rescoring of our own model on the contaminated ones (\cref{app:contamination}).
\end{itemize}

\section{Related Work}

Generative guardrails differ in what they moderate and in how the policy reaches them. Qwen3Guard~\cite{qwenguard}, Llama Guard~\cite{llamaguard} and ShieldGemma~\cite{shieldgemma} cover multiple safety domains and languages against fixed category sets. WildGuard~\cite{wildguard} targets adversarial prompts and jailbreaks, while PolyGuard~\cite{polyguard} emphasizes multilingual moderation. Other systems expose the policy more directly: YuFeng-XGuard~\cite{xguard} adds risk reasoning and a runtime policy interface, SingGuard~\cite{singguard} accepts the active policy at runtime across modalities, and Shieldstral-1.0-3B~\cite{shieldstral} scores text and images against natural-language policies. OpenGuardrails~\cite{openguardrails} packages a 15B judge as a deployable platform.

The reports for these models do not give results specifically for Russian prompt injection or Russian surface obfuscation, even when a model supports Russian. The 8B and 12B guards in \cref{tab:comparison} take 63 to 243\,ms in our benchmark run, compared with 14.3\,ms for HiveTraceGuard-Pro. Serving path matters as well as model size (\cref{sec:comparison}), so we treat the larger guards as accuracy references and include sub-1B variants where available.

\section{HiveTraceGuard-Pro Model}

\subsection{Threat Model}

\textbf{Prompt injection} attempts to override or ignore system instructions through instruction replacement or role manipulation. \textbf{Jailbreak attacks} preserve malicious intent while reframing a request through role-play, persuasion, hypothetical scenarios or other semantic transformations. \textbf{Surface obfuscation} changes the textual representation without changing the intended meaning. Examples include transliteration, keyboard-layout substitution, character replacement, random capitalization and symbol insertion. \Cref{tab:threatmap} lists eight training transforms and \cref{tab:peraug_full} lists eight evaluation transforms. Five occur in both sets. The evaluation-only transforms are \texttt{to\_lower}, \texttt{to\_upper} and \texttt{informal\_rewrite}.

HiveTraceGuard-Pro performs binary safety classification for user requests and assistant responses. Its chat template preserves \texttt{user} and \texttt{assistant} roles and selects the last eligible turn as the target. In the natural deployment path, the request guard judges the final \texttt{user} turn before generation and the response guard judges the final \texttt{assistant} turn before delivery. The response benchmarks in this report use a legacy standalone-reply path instead (\cref{app:contract}). They pass each reply as a standalone \texttt{user} turn, so they measure reply-content classification without exercising the template's assistant-target branch.

The threat model covers injected instructions that appear in the moderated turn, regardless of their origin. An instruction in a retrieved document that is never shown to the guard is out of scope. The evaluations target one turn per example and do not test behavior conditioned on a longer dialogue history. They also do not test adaptive attacks against the released weights. Every evaluated transform was fixed without access to model gradients.

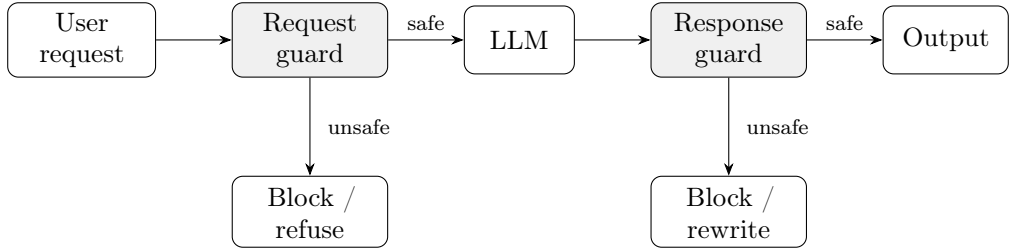
\begin{figure}[htbp]
\centering
\begin{tikzpicture}[node distance=8mm and 10mm]
  \node[fbox, text width=1.7cm] (user) {User\\request};
  \node[fguard, right=of user, text width=1.8cm] (rg) {Request\\guard};
  \node[fbox, right=of rg, text width=1.2cm] (llm) {LLM};
  \node[fguard, right=of llm, text width=1.8cm] (sg) {Response\\guard};
  \node[fbox, right=of sg, text width=1.4cm] (out) {Output};
  \node[fbox, below=13mm of rg, text width=1.8cm] (b1) {Block /\\refuse};
  \node[fbox, below=13mm of sg, text width=1.8cm] (b2) {Block /\\rewrite};
  \draw[-{Stealth[]}] (user) -- (rg);
  \draw[-{Stealth[]}] (rg) -- node[above, font=\scriptsize] {safe} (llm);
  \draw[-{Stealth[]}] (llm) -- (sg);
  \draw[-{Stealth[]}] (sg) -- node[above, font=\scriptsize] {safe} (out);
  \draw[-{Stealth[]}] (rg) -- node[midway, right=1mm, font=\scriptsize] {unsafe} (b1);
  \draw[-{Stealth[]}] (sg) -- node[midway, right=1mm, font=\scriptsize] {unsafe} (b2);
\end{tikzpicture}
\caption{Conceptual guard placement. In the natural deployment path, the request guard targets the final \texttt{user} turn and the response guard targets the final \texttt{assistant} turn. Reported response benchmarks instead use the legacy standalone-reply serialization of \cref{app:contract}. An unsafe verdict is a triage signal, and the surrounding application decides whether to block, refuse or rewrite.}
\label{fig:pipeline}
\end{figure}

\subsection{Model Architecture and Training Data}

HiveTraceGuard-Pro is built on Qwen3-0.6B and fine-tuned with LoRA\@. Its scoring rule returns one binary verdict without an explanation or category label. The generative formulation uses the causal language-model head already present in the base model. During training, only the verdict token and end-of-sequence marker contribute to the completion-only loss. \Cref{app:training} gives a configuration summary and the measured inference footprint.

The training corpus combines moderation datasets, translated resources, synthetic adversarial examples and internally generated robustness augmentations. It contains Russian and English examples in both dialogue roles. A labeled subset of the unsafe examples, 35\,013 rows under the 14 reported harm categories of \cref{tab:taxonomy} and 2\,948 under one additional policy category, is used for balancing and stratified evaluation, although the model predicts only a binary label. Where possible, each harmful example is paired with a semantically related benign example so that false positives and false negatives can be measured within the same topic. \Cref{tab:taxonomy} reports the training count for each presented category. \Cref{app:data} describes sources, licensing, deduplication and augmentation.

\section{Evaluation}

\subsection{Evaluation Methodology}
\label{sec:protocol}

\paragraph{Two cohorts.} The \emph{wider suite} contains thirty-five openly released guards. We score each guard through a model-specific adapter that preserves its published prompt format, decode settings and verdict parsing. Response items use the common standalone-reply construction described below. The harness covers nineteen benchmark groups drawn from forty-four datasets. Sixteen groups are public and three are our internal sets (\cref{app:suite}). We order the suite by the aggregate key, the geometric mean of the nineteen group scores. Seven groups have no benign class and therefore contribute recall alone. Their group scores do not measure over-blocking. \Cref{tab:groups} defines each group score. Cross-model comparisons within a cell use that cell's metric.

The \emph{head-to-head} cohort in \cref{tab:comparison} is a fifteen-guard subset of the suite: the eleven highest-scoring guards under the aggregate key and four widely deployed references below them. The cohort spans 0.5B to 15B parameters. All cells are our reruns rather than values copied from model reports. We checked the cohort tables against benchmark artifacts, the held-out tables against evaluation reports and the contamination results against a complete rerun of the audit. The head-to-head comparison has not been reproduced by a third party.

\paragraph{Two harnesses.} Cohort results and internal held-out results come from separate harnesses, so their values are not interchangeable even when they use the same rows. On augmented requests they agree to two decimals, with combined-F1 0.853 in the cohort harness and harm-F1 0.852 in the internal harness. Their error rates differ slightly, with FPR 0.132 versus 0.139 and FNR 0.128 versus 0.123. On the underlying counts these are 475 versus 500 false positives among the 3\,599 benign rows and 354 versus 339 misses among the 2\,756 harmful ones, and the paired row-level disagreement can be larger. Response-side unsafe-class F1 differs more, at 0.756 in the cohort harness and 0.671 in the internal harness. Their error rates also differ more, with FPR 0.026 versus 0.043 and FNR 0.164 versus 0.180. Only 6.5\% of the 945 response rows are positive, so the fifteen additional false positives and one further missed harmful reply in the internal harness reduce precision from 0.689 to 0.568. \Cref{tab:comparison} uses the cohort harness that scored all fifteen models. Comparisons among internal results use the internal harness. A model is graded only on rows for which it emits a usable verdict. \Cref{app:coverage} reports the per-model shortfalls.

\paragraph{Three F1 aggregations.} \emph{macro-F1} is the unweighted mean of safe-class and unsafe-class F1. \emph{harm-F1} is unsafe-class F1 alone. \emph{combined-F1} is the same unsafe-class F1 formula computed by the cohort harness over pooled safe and harmful slices. Harm-F1 and combined-F1 therefore differ in the harness and rows used, not in the formula. These quantities are not interchangeable. On the augmented \texttt{robustness-test}, for example, macro-F1 is 0.866 and harm-F1 is 0.852. We also report false-positive and false-negative rates to separate over-blocking from missed harm. Attack-only sets contain no benign class, so we report recall. Results on the confirmation set use 95\% Wilson intervals because every cell has fewer than 1\,000 examples (\cref{tab:confirm}). Other tables report point estimates. We do not report confidence intervals or multiplicity-corrected comparisons for cross-model differences or the suite ordering.

\paragraph{Evaluation sets.} We use four internal held-out sets (\cref{tab:evalsets}). None is a training split, and all four serve as references in the contamination audit. Three influenced accept or reject decisions during development and are reported as development sets. The fourth, \texttt{hivetrace/insecure-prompts}, was reserved and not queried until the released model was final (\cref{sec:confirm}). We also use public benchmarks where noted. In the Russian development sets, each harmful prompt is matched with a benign prompt from the same domain, such as Fascism versus History of fascism or Drugs versus Pharmacology.

\paragraph{Decision rule and decision path.} The scoring rule returns one verdict. We take the larger of the \texttt{safe} and \texttt{unsafe} logits, which is equivalent to a fixed 0.5 boundary under a two-way softmax. The boundary is not calibrated separately by language. The shipped path has no input normalizer, and reported labels are direct model outputs. Request-side evaluation matches the natural deployment path. Response-side evaluation does not. Every reported response result passes the reply standalone as a \texttt{user} turn, including the cohort comparison in \cref{tab:comparison}. A natural dialogue would pass the reply as an \texttt{assistant} turn and select a different branch of the chat template. Apart from the Aegis-2.0 check in \cref{app:contract}, that assistant-role path is not evaluated here. The primary internal augmented result is harm-F1 on the robustness set, which we call robustness-augmented harm-F1.

\begin{table}[htbp]
\centering
\small
\begin{adjustbox}{max width=\textwidth}
\begin{tabular}{@{}L{3.2cm}L{2.2cm}L{1.2cm}L{2.6cm}L{4.4cm}@{}}
\toprule
\textbf{Set} & \textbf{Role} & \textbf{Lang} & \textbf{Rows scored} & \textbf{Composition} \\
\midrule
\texttt{hivetrace/\allowbreak robustness-test}
  & development, primary
  & RU
  & 11\,879 \newline (10\,934 req / 945 resp)
  & paired safe/harm twins, requests crossed with clean and eight obfuscation transforms \\
RU-categories
  & development
  & RU
  & 220 per role per category
  & 17 topic-paired safe$\leftrightarrow$harm categories, request and response \\
\texttt{nvidia/\allowbreak Aegis-2.0}~\cite{aegis}
  & development, secondary
  & EN
  & 1\,964 req / 813 resp
  & externally authored, out-of-distribution English holdout, lower-confidence reference \\
\texttt{hivetrace/\allowbreak insecure-prompts}
  & \emph{reserved until final model}
  & RU
  & 156 req / 156 resp
  & every request an injection attempt (no benign class). Responses 85 harm / 71 safe \\
\bottomrule
\end{tabular}
\end{adjustbox}
\caption{The four internal held-out sets. The first three influenced development decisions. The fourth was reserved before development and evaluated once after the released model was final (\cref{sec:confirm}). Only \texttt{nvidia/Aegis-2.0} was authored outside our team. Aegis-2.0 response row counts differ between the internal and cohort harnesses, 813 versus 852, entirely because of benign rows. This table reports the internal count.}
\label{tab:evalsets}
\end{table}

\subsection{Comparison with Existing Guardrails}
\label{sec:comparison}

Results vary by domain. HiveTraceGuard-Pro has clean Russian robustness combined-F1 0.88, the highest value among the fifteen models and also the maximum in the wider suite. Its Russian prompt-injection recall is 0.999, with 4 misses among 4\,417 attacks. On English prompt injection it reaches 0.88, compared with 0.90 for YuFeng-XGuard-Reason-8B. On augmented requests it reaches 0.853 and on responses 0.756. Three guards score higher on the augmented column and three on the response column, where SingGuard-2B is level with ours at the table's precision. Across the full suite, three guards exceed the augmented-request score and five exceed the response score. Because the injection and S-Eval sets contain only attacks, their recall should be read together with benign over-blocking. Request-side FPR is 0.016 on clean inputs and 0.262 on \texttt{wrong\_layout}, the highest among the individual transforms. Only the internal harness breaks these errors down transform by transform (\cref{tab:peraug_full}).

\begin{figure}[H]
\centering
\figbox{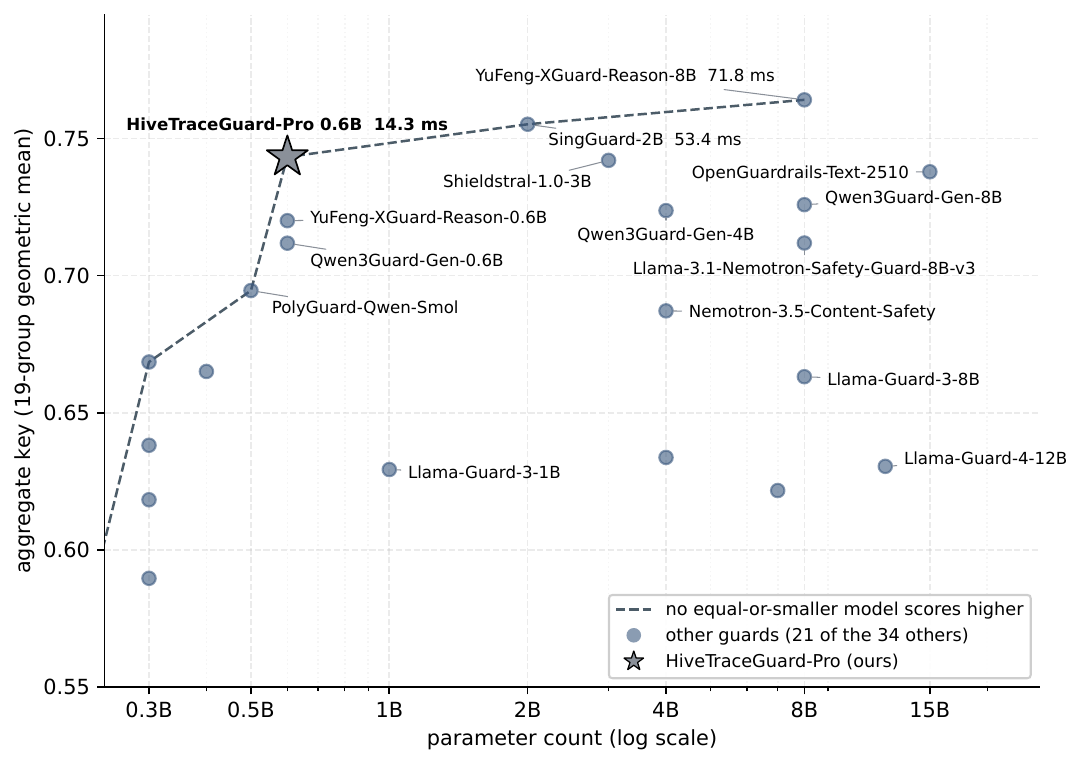}
\caption{\textbf{Aggregate key against parameter count.} The dashed line joins models for which no model of the same or smaller size has a higher aggregate key. Points show the 22 of 35 suite guards with an aggregate key above 0.55, and unnamed points are listed in \cref{tab:suite}. The two guards above HiveTraceGuard-Pro on the aggregate key use 2B and 8B parameters and have median latencies of 53.4 and 71.8\,ms in the same run, compared with 14.3\,ms here. Each guard was run once, so no uncertainty intervals are shown.}
\label{fig:cost}
\end{figure}

The 14.3\,ms median latency is the lowest value in the table, 0.41\,ms below Shieldstral-1.0-3B. The two models with higher aggregate keys take 53.35 and 71.79\,ms, and Qwen3Guard-Gen-8B takes 152.71\,ms. These values are comparable only within the same benchmark run. \Cref{fig:cost} plots the twenty-two guards with an aggregate key above 0.55.

The two other 0.6B guards in \cref{tab:comparison} have aggregate keys of 0.7200 for YuFeng-XGuard-Reason-0.6B and 0.7118 for Qwen3Guard-Gen-0.6B, compared with 0.7432 here. Relative to YuFeng-XGuard-Reason-0.6B, HiveTraceGuard-Pro has higher values on every Russian column, English prompt injection and Aegis-2.0 responses, and lower latency. It has lower values on both S-Eval columns, Aegis-2.0 prompts and ToxicChat. Relative to Qwen3Guard-Gen-0.6B, it has higher values on seven of ten score columns and lower latency, but lower values on both Aegis-2.0 columns and ToxicChat. Most of the fifteen models score higher on Aegis-2.0 prompts, Aegis-2.0 responses and ToxicChat. \Cref{sec:external} examines English performance on additional public benchmarks.

Latency in \cref{tab:comparison} depends on the serving configuration. The same HiveTraceGuard-Pro weights measure p50 32.75\,ms under our \texttt{-{}-enforce-eager} configuration and 14.3\,ms in the benchmark run. The models in the column come from that benchmark run, so within-run ratios are comparable, but absolute values are not portable. Runtime also depends on hardware, engine configuration, output length and serving implementation, not only parameter count. For example, a 1B guard takes 20.50\,ms while another 0.6B guard takes 37.18\,ms. HiveTraceGuard-Pro emits one token (\cref{app:contract}). The column therefore measures end-to-end harness latency and does not isolate model compute.

Every guard uses its published prompt format, decode settings and verdict strings through a per-model adapter. The data comparison is still asymmetric. Three of the nineteen groups are internal sets that we authored and used during development. Removing those groups drops HiveTraceGuard-Pro below two further guards on the key recomputed over the sixteen public groups (\cref{app:suite}). An unexpectedly low competitor cell may reflect its adapter or the benchmark as well as the model. We therefore interpret size comparisons primarily within the Qwen3Guard-Gen 0.6B/4B/8B and YuFeng-XGuard-Reason 0.6B/8B model families, and on the public Aegis-2.0 columns.

Llama-3.1-Nemotron-Safety-Guard-8B-v3 is flagged for evaluation leakage on Aegis-2.0. Its training data is partly derived from Aegis-2.0 and overlaps its holdout splits. We report the measured values without adjustment and mark the overlap. Our own model also has English-side Aegis-2.0 overlap, quantified in \cref{app:contamination}.

\begin{table}[htbp]
\centering
\scriptsize
\setlength{\tabcolsep}{3pt}
\renewcommand{\arraystretch}{1.15}
\captionsetup{singlelinecheck=false}
\begin{adjustbox}{max width=\textwidth}
\begin{tabular}{@{}llccccccccccc@{}}
\toprule
\textbf{Model} & \textbf{Params} & \makecell{Aegis-2.0\\combined-F1\\(prompt)} & \makecell{Aegis-2.0\\combined-F1\\(resp.)} & \makecell{Robust-clean\\combined-F1\\(FPR/FNR)} & \makecell{Robust-in\\combined-F1\\(FPR/FNR)} & \makecell{Robust-out\\combined-F1\\(FPR/FNR)} & \makecell{PI-EN\\recall} & \makecell{PI-RU\\recall} & \makecell{S-Eval\\base} & \makecell{S-Eval\\atk} & \makecell{Toxic\\Chat} & \makecell{p50\\(ms)} \\
\midrule
YuFeng-XGuard-Reason-8B~\cite{xguard} & 8B & 0.85 & 0.78 & \makecell{0.81\\{\scriptsize(.03/.10)}} & \makecell{0.86\\{\scriptsize(.03/.21)}} & \makecell{0.71\\{\scriptsize(.00/.43)}} & \textbf{0.90} & 0.909 & \textbf{0.81} & \textbf{0.95} & 0.55 & 71.79 \\
SingGuard-2B~\cite{singguard} & 2B & 0.84 & 0.82 & \makecell{0.87\\{\scriptsize(.01/.12)}} & \makecell{0.84\\{\scriptsize(.04/.24)}} & \makecell{0.76\\{\scriptsize(.01/.28)}} & 0.76 & 0.742 & 0.68 & 0.63 & 0.54 & 53.35 \\
\textbf{HiveTraceGuard-Pro (ours)} & 0.6B & 0.82$^\dagger$ & 0.80 & \makecell{\textbf{0.88}\\{\scriptsize(.02/.05)}} & \makecell{0.85\\{\scriptsize(.13/.13)}} & \makecell{0.76\\{\scriptsize(.03/.16)}} & 0.88$^\dagger$ & \textbf{0.999}$^\dagger$ & 0.71 & 0.80$^\dagger$ & 0.51$^\dagger$ & \textbf{14.30} \\
Shieldstral-1.0-3B~\cite{shieldstral} & 3B & 0.85 & 0.81 & \makecell{0.72\\{\scriptsize(.04/.13)}} & \makecell{0.82\\{\scriptsize(.08/.24)}} & \makecell{0.71\\{\scriptsize(.01/.34)}} & 0.74 & 0.836 & 0.73 & 0.61 & \textbf{0.61} & 14.71 \\
OpenGuardrails-Text-2510~\cite{openguardrails} & 15B & 0.82 & \textbf{0.87} & \makecell{0.77\\{\scriptsize(.01/.32)}} & \makecell{0.73\\{\scriptsize(.03/.40)}} & \makecell{0.75\\{\scriptsize(.00/.38)}} & 0.72 & 0.687 & 0.52 & 0.49 & 0.56 & 63.22 \\
Qwen3Guard-Gen-8B~\cite{qwenguard} & 8B & 0.86 & 0.84 & \makecell{0.80\\{\scriptsize(.03/.08)}} & \makecell{\textbf{0.89}\\{\scriptsize(.05/.15)}} & \makecell{\textbf{0.82}\\{\scriptsize(.00/.26)}} & 0.83 & 0.904 & 0.72 & 0.68 & 0.58 & 152.71 \\
Qwen3Guard-Gen-4B~\cite{qwenguard} & 4B & 0.86 & 0.83 & \makecell{0.76\\{\scriptsize(.04/.09)}} & \makecell{0.88\\{\scriptsize(.04/.18)}} & \makecell{0.77\\{\scriptsize(.01/.33)}} & 0.83 & 0.908 & 0.72 & 0.65 & 0.58 & 96.92 \\
YuFeng-XGuard-Reason-0.6B~\cite{xguard} & 0.6B & 0.86 & 0.77 & \makecell{0.78\\{\scriptsize(.03/.12)}} & \makecell{0.78\\{\scriptsize(.05/.31)}} & \makecell{0.67\\{\scriptsize(.00/.49)}} & 0.87 & 0.918 & 0.79 & \textbf{0.95} & 0.53 & 16.85 \\
Llama-3.1-Nemotron-Safety-Guard-8B-v3 & 8B & \textbf{0.87}$^{\S}$ & 0.86$^{\S}$ & \makecell{0.68\\{\scriptsize(.04/.26)}} & \makecell{0.71\\{\scriptsize(.07/.39)}} & \makecell{0.72\\{\scriptsize(.00/.42)}} & 0.85 & 0.870 & 0.67 & 0.60 & 0.51 & 243.28 \\
Qwen3Guard-Gen-0.6B~\cite{qwenguard} & 0.6B & 0.85 & 0.82 & \makecell{0.63\\{\scriptsize(.07/.14)}} & \makecell{0.81\\{\scriptsize(.12/.21)}} & \makecell{0.70\\{\scriptsize(.01/.39)}} & 0.73 & 0.894 & 0.70 & 0.61 & 0.55 & 37.18 \\
PolyGuard-Qwen-Smol~\cite{polyguard} & 0.5B & 0.85 & 0.73 & \makecell{0.67\\{\scriptsize(.04/.22)}} & \makecell{0.70\\{\scriptsize(.08/.41)}} & \makecell{0.51\\{\scriptsize(.05/.43)}} & 0.79 & 0.861 & 0.65 & 0.59 & 0.44 & 77.36 \\
Nemotron-3.5-Content-Safety & 4B & 0.86 & 0.85 & \makecell{0.60\\{\scriptsize(.09/.10)}} & \makecell{0.85\\{\scriptsize(.11/.16)}} & \makecell{0.67\\{\scriptsize(.00/.48)}} & 0.83 & 0.930 & 0.73 & 0.65 & 0.49 & 54.90 \\
Llama-Guard-3-8B & 8B & 0.76 & 0.64 & \makecell{0.61\\{\scriptsize(.02/.43)}} & \makecell{0.72\\{\scriptsize(.08/.38)}} & \makecell{0.47\\{\scriptsize(.01/.66)}} & 0.71 & 0.603 & 0.43 & 0.42 & 0.30 & 63.16 \\
Llama-Guard-4-12B & 12B & 0.71 & 0.63 & \makecell{0.49\\{\scriptsize(.07/.38)}} & \makecell{0.78\\{\scriptsize(.24/.16)}} & \makecell{0.30\\{\scriptsize(.05/.69)}} & 0.66 & 0.529 & 0.36 & 0.45 & 0.26 & 86.19 \\
Llama-Guard-3-1B & 1B & 0.73 & 0.61 & \makecell{0.48\\{\scriptsize(.09/.32)}} & \makecell{0.77\\{\scriptsize(.13/.27)}} & \makecell{0.24\\{\scriptsize(.18/.51)}} & 0.68 & 0.635 & 0.49 & 0.59 & 0.18 & 20.50 \\
\bottomrule
\end{tabular}
\end{adjustbox}
\caption{\textbf{Head-to-head comparison.} Fifteen suite guards from 0.5B to 15B parameters, scored by us in one benchmark run under \cref{sec:protocol} and ordered by the same nineteen-group aggregate key. Bold marks the best displayed value in each column, so two cells that round to the same figure are both bold. Counts of how many guards score higher, here and in \cref{sec:limitations}, compare the underlying values. Each Aegis-2.0 column is unsafe-class F1 over pooled safe and harmful slices, with prompts and responses scored separately. \emph{Robust-clean}, \emph{Robust-in} and \emph{Robust-out} are the clean-request, augmented-request and response portions of \texttt{robustness-test}. Each is reported as combined-F1 with FPR/FNR below. PI-EN and PI-RU contain only attacks, so they report recall and their FNR is one minus recall. They are translations of the same 4\,417 upstream items, not independent benchmarks. S-Eval base and attack report recall~\cite{seval}. ToxicChat reports unsafe-class F1~\cite{toxicchat}. The p50 column is median wall-clock latency with one request in flight (\cref{tab:training}). Each model uses a per-model adapter for its published prompt, decode settings and verdict format. The response side uses the common standalone-reply construction. $\dagger$: training overlap is at least 27.1\% for PI-RU, 10.0\% for S-Eval attack, 21.3\% for the Aegis-2.0 prompt set and 3.2\% for ToxicChat (\cref{app:contamination}). The contamination scan is monolingual, so the low measured overlap on PI-EN does not establish training independence. $\S$: evaluation-leak flag, described in \cref{sec:comparison}. Response columns use the legacy standalone-reply protocol (\cref{app:contract}).}
\label{tab:comparison}
\end{table}

\subsection{Internal Held-Out Evaluation}
\label{sec:internal}

On clean classification, macro-F1 is 0.952 on the synthetic-clean subset, the 450 benign and 345 harmful paired examples to which the obfuscation transforms are applied, and 0.937 on RU-categories. Harm-F1 is 0.946 on the synthetic-clean subset. AUROC is 0.997 over the 4\,579 clean request rows and 0.985 on RU-categories.

The primary augmented result is harm-F1 0.852 (\cref{tab:internal_results}). Macro-F1 is 0.898 when clean and augmented conditions are pooled. Over the combined request and response \texttt{robustness-test} ($n=11\,879$), the model correctly labels 8\,110 benign turns and flags 607 benign turns. It detects 2\,798 harmful turns and misses 364. \Cref{sec:errors} breaks these errors down by condition and category. Across 32 reported RU-categories cells, unsafe-class F1 ranges from 0.843 to 1.000, with 25 cells at or above 0.90 and a minimum request-side value of 0.875 (\cref{tab:percat_numeric}).

We also evaluate a prompted baseline under the same scoring rule. The untuned Qwen3-0.6B backbone receives an English guard prompt and uses the same constrained verdict-token argmax. It misses most harmful requests, with harm-FNR 0.87 on Aegis-2.0 and 0.93 on the robustness set, and over-blocks benign responses at FPR 0.49--0.77. The released model's harm-FNR is 0.206 and 0.114 on those request sets, and its benign-response FPR is 0.193 and 0.043. A simple Russian prompt that requests JSON output fails at the output-format stage, producing RU-categories FPR 0.169 and FNR 0.770. This comparison changes the prompt and scoring setup as well as the weights, so it does not isolate the effect of fine-tuning.

\begin{table}[htbp]
\centering
\footnotesize
\begin{tabular}{@{}llcccrr@{}}
\toprule
\textbf{Held-out set} & \textbf{Condition} & \textbf{macro-F1} & \textbf{FPR} & \textbf{FNR} & \textbf{safe $n$} & \textbf{harm $n$} \\
\midrule
robustness-test & synthetic-clean subset & 0.952 & 0.053 & 0.041 & 450 & 345 \\
robustness-test & robust (augmented) & 0.866 & 0.139 & 0.123 & 3\,599 & 2\,756 \\
robustness-test & clean requests & 0.939 & 0.016 & 0.041 & 4\,234 & 345 \\
robustness-test & requests only (clean+aug) & 0.898 & 0.073 & 0.114 & 7\,833 & 3\,101 \\
robustness-test & responses & 0.821 & 0.043 & 0.180 & 884 & 61 \\
robustness-test & overall (clean+aug, both roles) & 0.898 & 0.070 & 0.115 & 8\,717 & 3\,162 \\
RU-categories & request + response & 0.937 & 0.082 & 0.044 & 3\,740 & 3\,740 \\
Aegis-2.0 (EN) & request & 0.815 & 0.159 & 0.206 & 905 & 1\,059 \\
Aegis-2.0 (EN) & response & 0.807 & 0.193 & 0.193 & 419 & 394 \\
Aegis-2.0 (EN) & request + response & 0.813 & 0.170 & 0.202 & 1\,324 & 1\,453 \\
\bottomrule
\end{tabular}
\caption{The three development sets under the internal harness. \Cref{tab:confirm} reports the reserved fourth set. Counts come from the run's classification reports. The augmented-only result is macro-F1 0.866 and harm-F1 0.852. The Aegis-2.0 values differ from \cref{tab:comparison}. Both harnesses score the same 1\,964 requests and give similar rates, FPR 0.159 versus 0.152 and FNR 0.206 versus 0.211.}
\label{tab:internal_results}
\end{table}

\subsection{Reserved Confirmation Set}
\label{sec:confirm}

The three development sets influenced accept or reject decisions, and we did not record how often each set was queried. Their results may therefore be optimistic. We reserved a fourth set, \texttt{hivetrace/insecure-prompts}, before development. It was part of the decontamination reference but not of the residual contamination scan (\cref{app:contamination}), and it was not used for model selection. We evaluated the frozen release on it once with the same contract and internal harness used for \cref{tab:internal_results}, and made no changes after observing the result. Because this set differs in composition from the development sets, it is an out-of-development check rather than a direct estimate of selection bias.

The set contains 156 Russian dialogues. Every user turn is an injection attempt: 120 jailbreaks, 22 prompt injections, 13 harmful-content requests and 1 data-extraction request. Each dialogue has an independently labeled assistant reply, giving 85 harmful and 71 safe replies. The request block has no benign class, so it has no measurable FPR or macro-F1. Its harm-F1 is determined entirely by recall because unsafe-class precision is fixed at 1.0. An always-unsafe classifier would therefore score perfectly on the request block.

The pooled request-plus-response result ($n=312$, macro-F1 0.655) mixes this one-class request block with the response block and is not used as a summary result. The response block is the only part that supports macro-F1. In the same run, rescoring the three development sets reproduced all 18 metric aggregates to ten decimal places. Responses use the same legacy standalone-reply path as the other reported response results, so the model does not see the associated injection prompt.

\begin{table}[htbp]
\centering
\small
\begin{adjustbox}{max width=\textwidth}
\begin{tabular}{@{}lrcccc@{}}
\toprule
\textbf{Slice} & \textbf{$n$} & \textbf{macro-F1} & \textbf{harm-F1} & \textbf{FPR (95\% CI)} & \textbf{FNR (95\% CI)} \\
\midrule
\multicolumn{6}{@{}l}{\emph{insecure-prompts (reserved)}} \\
Requests, all \texttt{injection} & 156 & n/a & 0.953 & n/a & 0.090 (0.054--0.145) \\
Responses, safe slice & 71 & n/a & n/a & 0.563 (0.448--0.673) & n/a \\
Responses, harm slice & 85 & n/a & 0.874 & n/a & 0.224 (0.148--0.323) \\
Responses, pooled & 156 & 0.602 & 0.691 & 0.563 (0.448--0.673) & 0.224 (0.148--0.323) \\
\midrule
\multicolumn{6}{@{}l}{\emph{Development held-out sets, response side, same scoring run}} \\
robustness-test & 945 & 0.821 & 0.671 & 0.043 (0.031--0.058) & 0.180 (0.104--0.295) \\
Aegis-2.0 (EN) & 813 & 0.807 & 0.802 & 0.193 (0.158--0.234) & 0.193 (0.157--0.235) \\
\bottomrule
\end{tabular}
\end{adjustbox}
\caption{The released model on the reserved set, together with two development response sets from the same scoring run. Intervals are 95\% Wilson intervals for the underlying rates. An \texttt{n/a} marks undefined quantities. The request block contains only injection attempts, so FPR and macro-F1 are undefined. Each single-label response slice supports only its corresponding error rate. On the 85-row harmful-response slice, precision is fixed at 1.000 because no benign rows are present. Harm-F1 0.874 is therefore a transformation of recall, with 66 of 85 rows predicted unsafe.}
\label{tab:confirm}
\end{table}

On this attack-only request block, harm-F1 is 0.953 and FNR is 0.090, compared with request-side FNR 0.114 on \texttt{robustness-test} and 0.206 on Aegis-2.0. The model detects all 22 prompt injections, 108 of 120 jailbreaks and 12 of 14 harmful-content or data-extraction requests. The 14 misses are 12 jailbreaks and 2 requests from the remaining categories. Because the set has no benign requests, this result does not establish better overall request moderation than on the development sets.

Response over-blocking is much higher on the reserved set. The model flags 40 of 71 safe replies, giving FPR 0.563, compared with 0.043 and 0.193 on the development sets. The 95\% Wilson interval is wide, but its lower bound of 0.448 is above the upper bounds for both development results. Errors occur in both response types: 22 of 40 safe refusals and 18 of 31 safe compliances are flagged. The replies are in-character role-play continuations of jailbreak attempts and are labeled safe when they contain no harmful content. This narrow register is a plausible source of distribution shift. The set is small, entirely Russian and evaluated with one model seed, so it should be read as a stress test rather than a replacement for the development sets.

\subsection{Error Analysis}
\label{sec:errors}

On the Russian robustness set, 607 of 971 errors are benign turns flagged as unsafe. Most occur after augmentation: 500 errors among 3\,599 augmented benign requests, for FPR 0.139. The remaining false positives are 69 among 4\,234 clean requests and 38 among 884 benign responses. \Cref{app:coverage} compares this error profile with the rest of the suite.

Strong surface obfuscation remains difficult. The held-out evaluation builds its \texttt{wrong\_layout} transform with a cipher that leaves five Cyrillic letters unchanged (U+0431, U+0436, U+0445, U+044A, U+044E) and maps yo (U+0451) to a backtick. An earlier training-time cipher mapped those characters to punctuation, so the pre-fix model was evaluated on substitutions absent from training. On the pre-fix build that transform carried a request-side harm-FNR of 0.675. Aligning the ciphers, then rebalancing the obfuscation augmentation (\cref{app:ablation}), brought the released build to 0.125 at the same model size. Because the builds differ in more than the cipher, this comparison establishes the direction of change but is not a controlled ablation (\cref{tab:peraug_full}).

\begin{figure}[htbp]
\centering
\figbox{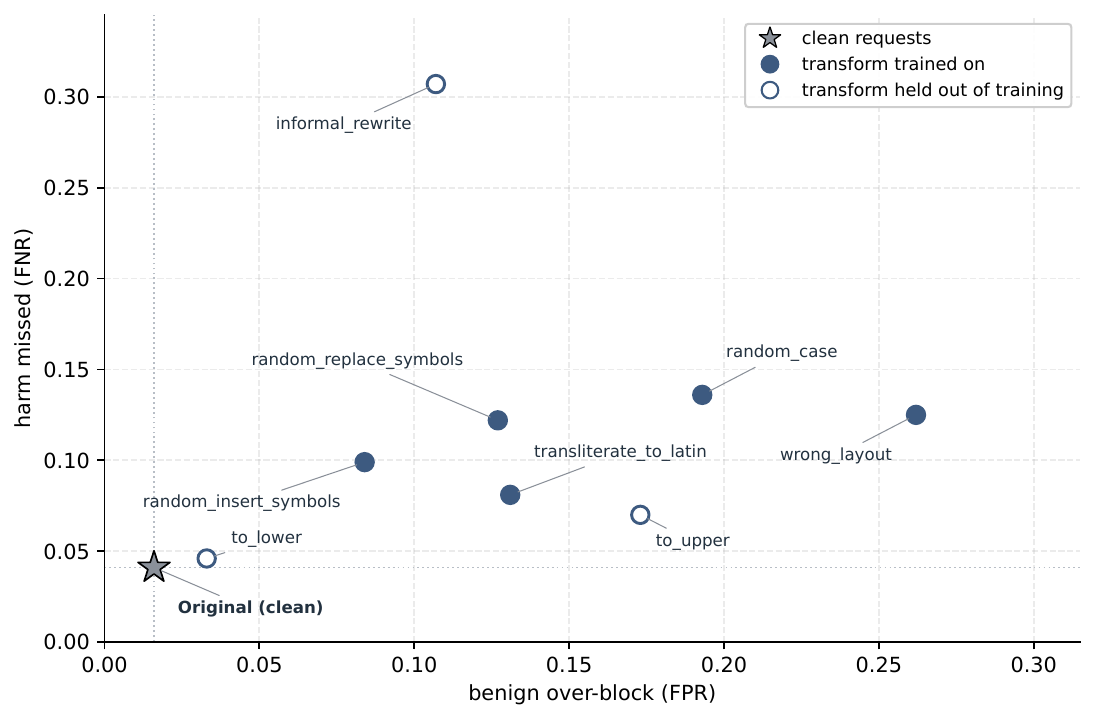}
\caption{Request-side FPR and FNR for each condition in \cref{tab:peraug_full}. The clean point uses 4\,234 benign rows, compared with 449 or 450 for each transform. Every transform worsens at least one error rate relative to clean requests. The \texttt{wrong\_layout} transform has the highest FPR (0.262), and the held-out \texttt{informal\_rewrite} has the highest FNR (0.307). Filled points denote transforms used in training, and hollow points denote evaluation-only transforms.}
\label{fig:obfusc}
\end{figure}

The remaining errors occur on both labels. Harmful requests are missed most often after the held-out \texttt{informal\_rewrite} transform, with harm-FNR 0.307, and after randomized case, with FNR 0.136. \Cref{fig:obfusc} shows all conditions together.

Across the additive rounds in \cref{app:ablation}, improvements on one obfuscation error rate repeatedly worsened the other. None of the tested data mixtures removed this coupling. Because all experiments keep the model at 0.6B, they do not distinguish a data limitation from a capacity limitation.

The largest response-side false-positive rates occur on benign discussion of harm-adjacent topics. Response-side FPR is 0.373 for cybercrime, 0.373 for non-violent crime and 0.291 for weapons, the three highest reported categories.

The three RU-categories topic pairs without a dedicated training class (\cref{tab:taxonomy}) test transfer rather than in-distribution performance. Military conflict has request-side harm-F1 0.875, the request-side minimum, and copyright infringement has 0.885. Both are among the four lowest request-side cells. All three pairs carry request-side FPR between 0.127 and 0.145. We do not rank them against one another at $n=220$ per cell.

\subsection{External Benchmark Generalization}
\label{sec:external}

Because the preceding results rely heavily on our held-out distribution, we also evaluate the released model on ten public benchmarks against four other guards (\cref{tab:external_bench}). Every cell is our measurement. Two competitors are also 0.6B, including the closest of the four below HiveTraceGuard-Pro on the aggregate key. Five are English benchmarks and five are multilingual slices.

Across these ten benchmarks, HiveTraceGuard-Pro has the highest value in three rows, the lowest in two and is within 1.5 points of the highest value in three. Its recall is 95.2 on the Russian Aya red-teaming slice and 91.7 on the English slice. Its unsafe-class F1 on XSafety English is 62.1. Absolute performance on XSafety is low across the suite. Four guards have FPR at or above 0.95, and among the other 31 the highest score is 67.5. Three of the four post larger nominal scores, 96.0, 73.0 and 71.7, at FPR 1.000, 0.975 and 1.000, and the fourth scores 56.2 at FPR 1.000.

The largest gaps are all on English benchmarks. SimpleSafetyTests recall is 91.0 compared with 99.0 for all four competitors. HarmBench response F1 is 82.7 compared with 86.3 for WildGuard. OpenAI Moderation F1 is 71.4 compared with 79.0 for Llama-Guard-3-8B, and ToxicChat F1 is 50.6 compared with 54.9 for Qwen3Guard-Gen-0.6B. The ToxicChat cell carries FNR 0.580 in the cohort harness. The English deficit is not uniform across paired datasets. On four of seven English/Russian pairs in the suite, English recall is at least as high as Russian recall, and the model posts the highest score of the five guards in \cref{tab:external_bench} on both Aya slices. These results point to benchmark-specific weaknesses rather than a single language-wide shift. \Cref{tab:groups} reports all nineteen group scores.

\begin{table}[htbp]
\centering
\footnotesize
\begin{adjustbox}{max width=\textwidth}
\begin{tabular}{@{}llccccc@{}}
\toprule
\textbf{Benchmark} & \textbf{Metric} & \makecell{HiveTraceGuard-Pro\\(ours, 0.6B)} & \makecell{YuFeng-XGuard-\\Reason-0.6B} & \makecell{Qwen3Guard-\\Gen-0.6B} & \makecell{WildGuard\\7B} & \makecell{Llama-Guard-3\\8B} \\
\midrule
Aya Red-teaming (RU)~\cite{ayart} & recall & \textbf{95.2} & 90.6 & 92.6 & 73.6 & 69.6 \\
Aya Red-teaming (EN)~\cite{ayart} & recall & \textbf{91.7} & 85.0 & 90.7 & 90.9 & 62.3 \\
XSafety (EN)~\cite{xsafety} & unsafe-class F1 & \textbf{62.1} & 60.6 & 60.7 & 61.4 & 39.8 \\
RTP-LX (RU, request)~\cite{rtplx} & unsafe-class F1 & 87.5 & \textbf{88.8} & 86.3 & 64.5 & 48.0 \\
RTP-LX (EN, request)~\cite{rtplx} & unsafe-class F1 & 93.0 & 91.1 & 91.2 & \textbf{94.5} & 48.8 \\
BeaverTails (responses)~\cite{beavertails} & unsafe-class F1 & 85.5 & 83.6 & \textbf{86.3} & 84.1 & 67.8 \\
HarmBench (responses)~\cite{harmbench} & unsafe-class F1 & 82.7 & 85.5 & 84.8 & \textbf{86.3} & 84.5 \\
OpenAI Moderation~\cite{openaimod} & unsafe-class F1 & 71.4 & 71.0 & 66.1 & 72.7 & \textbf{79.0} \\
ToxicChat~\cite{toxicchat} & unsafe-class F1 & 50.6 & 53.0 & \textbf{54.9} & 43.5 & 29.9 \\
SimpleSafetyTests~\cite{sst} & recall & 91.0 & \textbf{99.0} & \textbf{99.0} & \textbf{99.0} & \textbf{99.0} \\
\bottomrule
\end{tabular}
\end{adjustbox}
\caption{Public safety benchmarks ($\times100$), measured by us in one benchmark run. Each row names its metric. Rows with both classes report unsafe-class F1. Attack-only rows report recall, which is not directly comparable with F1. Bold marks the best value per row. The first five are multilingual slices of Aya Red-teaming, RTP-LX and XSafety (\cref{sec:external}). The YuFeng-XGuard-Reason-0.6B, Qwen3Guard-Gen-0.6B and Llama-Guard-3-8B ToxicChat values repeat the same runs used in \cref{tab:comparison}. Competitors are English-native or multilingual. HiveTraceGuard-Pro also has direct English supervision, with 187\,215 of 464\,092 training rows in English (40.3\%, \cref{tab:summary}).}
\label{tab:external_bench}
\end{table}

\section{Intended Use and Scope}
\label{sec:scope}

\paragraph{Intended use.} HiveTraceGuard-Pro is a request and response filter for LLM applications, positioned as in \cref{fig:pipeline}. We trained it on Russian and English, and evaluated it most extensively on Russian. Its binary verdict is an input to a moderation policy, not a complete policy decision. Response numbers come from the legacy standalone-reply path of \cref{app:contract}, not from the natural assistant-role path.

\paragraph{Out of scope.} The model is not validated for languages other than Russian and English. Other deployments require domain-specific evaluation. It does not return categories, explanations or policy reasoning. Applications that need those outputs require a separate classifier or policy layer. The model judges one final turn and is not evaluated as a monitor for multi-step agent trajectories or tool calls. It is text-only and does not provide a legal or compliance determination.

\paragraph{Operating point.} The decision rule is the argmax boundary of \cref{sec:protocol}. AUROC provides a threshold-independent comparison: 0.985 on RU-categories and 0.973 on pooled \texttt{robustness-test} rows, compared with 0.888 on Aegis-2.0. The English shortfall therefore reflects ranking quality as well as threshold placement. A global threshold sweep over the three development sets finds $\tau=0.532$ and changes mean macro-F1 by less than 0.001. Even a separate oracle threshold for each set improves macro-F1 by at most 0.003. On the \texttt{robustness-test} response slice the same sweep improves macro-F1 by 0.061. Operators serving substantial English traffic should calibrate on their own distribution.

\paragraph{Prompt handling.} The shipped tokenizer call applies no automatic truncation. The 2048-token value in \cref{tab:training} is a training limit, not an inference limit, and forward passes were verified through the model's 40\,960-position range. Behavior beyond that range is unspecified. In one exploratory probe, a harmful request padded to roughly 80\,000 filler tokens was labeled safe (\cref{app:contract}). Because the policy block opens the prompt and the target turn and its marker close it, neither end may be cut. Any application-side length cap must shorten the conversation between them (\cref{app:contract}).

\paragraph{Failure behavior.} If the guard is unavailable, a fail-open pipeline permits all traffic during the outage, while a fail-closed pipeline blocks all traffic. This operational choice is separate from the model's normal classification errors. At the shipped operating point on the Russian robustness set, the guard flags 7.0\% of benign turns and misses 11.5\% of harmful turns (\cref{sec:internal}). Individual slices vary substantially, from FPR 0.016 on clean requests to 0.563 on the reserved-set responses. Operators should estimate both error costs on data that resemble their own traffic (\cref{sec:limitations,sec:confirm}).

\section{Limitations}
\label{sec:limitations}

\begin{figure}[htbp]
\centering
\figbox{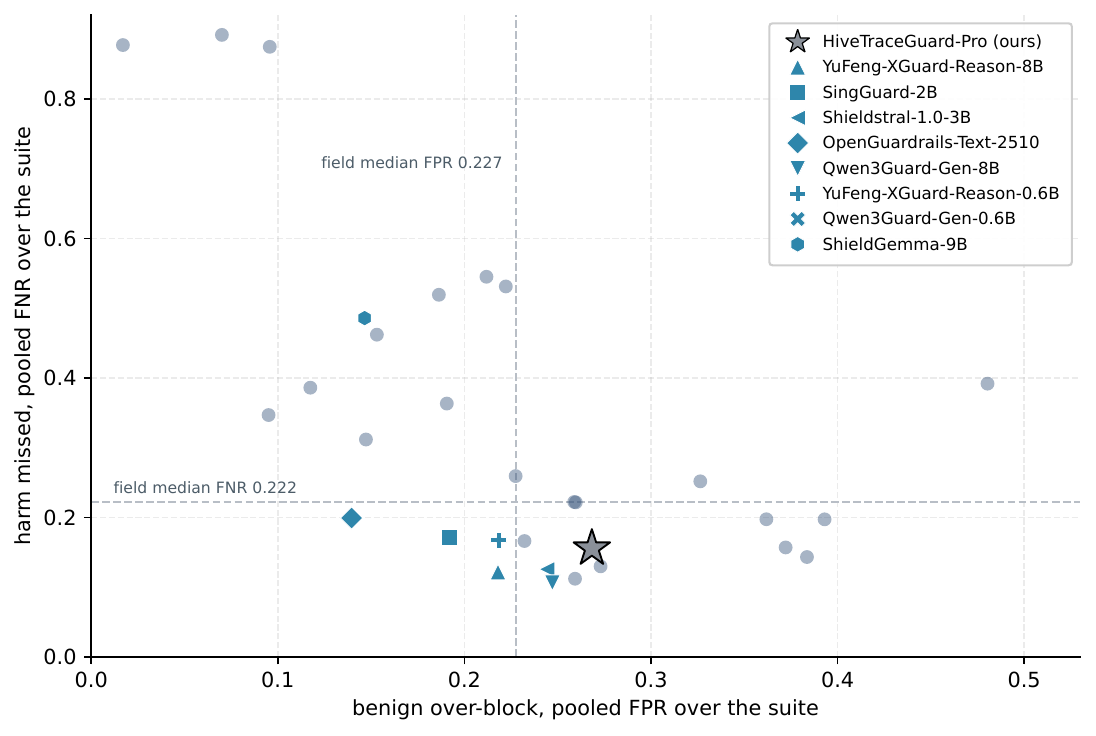}
\caption{Pooled FPR and FNR for each guard across the nineteen suite groups. Three injection-focused detectors fall outside the plotted range, at FPR 0.776, FPR 1.000 and FNR 0.989, and the medians include them. HiveTraceGuard-Pro has FPR 0.268 and FNR 0.156, compared with field medians of 0.227 and 0.222. Labels identify nearby guards on the aggregate key and selected reference models. Qwen3Guard-Gen-0.6B lies under the HiveTraceGuard-Pro marker because both have FPR 0.268 and their FNR differs by 0.004. The aggregate key weights groups equally, whereas this figure pools rows across sets of different sizes.}
\label{fig:fielderr}
\end{figure}

\textbf{Over-blocking is more common than missed harm.} Pooled over the suite, FPR is 0.268 and FNR is 0.156. Eight of the thirty-five guards have lower FNR, while twenty-five have lower FPR (\cref{fig:fielderr}). Averaging by benchmark group gives FPR 0.254 over the twelve groups with a benign class and FNR 0.184 over all nineteen groups. The same ordering holds under both aggregations. \Cref{app:coverage} identifies the benign-stress sets with the highest FPR\@.

\textbf{Response side.} Clean response harm-F1 is 0.671 compared with 0.889 on requests, mainly because of lower precision. Response-side FPR is 0.563 on the reserved set and 0.043 on \texttt{robustness-test} (\cref{sec:confirm}), so the development-set FPR does not transfer to the reserved response distribution. All reported response results use the legacy standalone-reply serialization with role \texttt{user}. Apart from the sensitivity check in \cref{app:contract}, they do not evaluate the natural \texttt{assistant}-role branch of the chat template.

\textbf{Obfuscation.} Across six additive data rounds and one change of training objective, improvements in recall on obfuscated harmful requests repeatedly increased over-blocking of obfuscated benign requests, or the reverse (\cref{sec:errors,app:ablation}). The tested changes did not remove this trade-off.

\textbf{Prompt length.} The tokenizer does not truncate automatically, and behavior beyond the verified 40\,960-position range is unspecified (\cref{sec:scope}). Deployments should therefore enforce a validated context limit while preserving both the policy prefix and the target suffix (\cref{app:contract}).

\textbf{Scope of the summary results.} The released robustness-augmented harm-F1 of 0.852 is above the three-seed mean of $0.844 \pm 0.011$. On the response robustness column it reaches combined-F1 0.76 against 0.82 for an 8B guard, with three cohort guards ahead, and it leads the clean-request column. Recall on attack-only sets measures detection only and does not include benign over-blocking. One global argmax boundary is used for both languages. Two suite cells are too small for reliable ranking, at $n=16$ and $n=61$. The RTP-LX English response FPR is based on 26 benign examples within a 999-row cell. Latency is measured with one request in flight and does not estimate concurrent throughput. Several orderings in this report turn on differences of a few thousandths, among them the response robustness column and the false-positive ranking above. Each guard was run once, so these are point estimates without intervals.

\section{Conclusion}

We presented HiveTraceGuard-Pro, a 0.6B generative guardrail for request and response moderation in Russian and English. We release the weights on Hugging Face under Apache-2.0. The evaluation compares thirty-five guards in one harness on nineteen benchmark groups, includes a confirmation set reserved before development and evaluated once on the frozen release (\cref{sec:confirm}), and audits overlap between the scored benchmarks and the training corpus (\cref{app:contamination}).

The model's aggregate key is 0.7432, behind an 8B guard at 0.7641 and a 2B guard at 0.7552. Restricted to the sixteen public groups, its key is 0.7153, with four of the thirty-four other suite guards above it (\cref{app:suite}). In the fifteen-model comparison it has the highest score on the clean Russian robustness set, which we authored and used during development. It also has the highest Russian prompt-injection recall and reaches 0.88 on the English prompt-injection set, compared with 0.90 for YuFeng-XGuard-Reason-8B. The Russian set contains only attacks and overlaps the training corpus by at least 27.1\%, and the English set translates the same items, so its low measured overlap does not establish training independence (\cref{app:contamination}). Three guards score higher on augmented requests and three on the legacy response evaluation.

HiveTraceGuard-Pro has the lowest median latency in the comparison, with a margin of 0.41\,ms over the next model, although absolute latency depends on the serving setup (\cref{sec:comparison}). Unsafe-class F1 is at least 0.90 on 25 of 32 reported category cells (\cref{tab:percat_numeric}).

The model's strongest results are on our in-domain Russian tests. Performance is weaker under strong obfuscation, on English general-moderation benchmarks and in response-side precision.

\section{Model Availability}
\label{sec:availability}

\begin{sloppypar}
HiveTraceGuard-Pro is released on Hugging Face under Apache-2.0 at \url{https://huggingface.co/hivetrace/HiveTraceGuard-Pro} as a single merged model. The training corpus, evaluation datasets and evaluation code remain internal. \Cref{app:contract} specifies prompt serialization, verdict-token ids and truncation behavior, and includes code that reproduces one reported result. The shipped \texttt{generation\_config.json} uses greedy one-token decoding with \texttt{max\_new\_tokens} 1 and \texttt{do\_sample} false. The README also documents caller-side constrained scoring with \texttt{allowed\_token\_ids=[18675,38157]} and a two-way softmax over the verdict logits.
\end{sloppypar}

The 0.6B model's weights occupy 1.11\,GiB, and allocator peak is 1.92\,GiB at a 2048-token context (\cref{tab:training}), or 1.39\,GiB with logits restricted to the final position (\cref{app:training}). In the comparison run it is about $10.7\times$ faster than Qwen3Guard-Gen-8B at 152.7\,ms. This ratio applies only to that run.

\appendix

\section{Dataset Construction}
\label{app:data}

\subsection{Sources and Admission}

The training corpus combines public safety datasets, translated moderation resources, internally curated Russian datasets and synthetic adversarial examples from the six provenance families in \cref{tab:provenance}. Public sources are filtered by license at the family level rather than source by source. Rows are admitted from existing label, category or score columns, or through a whole-source decision. We do not use regex, keyword or length filters to infer the safety label.

Because labels are inherited column-wise instead of re-annotated by multiple human raters, the corpus carries no inter-annotator-agreement statistic of its own. Peer corpora that do re-annotate report such measures (WildGuard~\cite{wildguard} Fleiss' kappa 0.55--0.72, Aegis~\cite{aegis} approximately 74\% agreement, PolyGuard~\cite{polyguard} Krippendorff's alpha 0.46 and 0.94), whereas our label reliability rests on the upstream sources' own labeling.

The corpus draws on 19\,551 distinct source tags. The only per-source breakdown available to us is an earlier assembly snapshot ($n=424\,040$), in which the largest single source is an in-house synthetic obfuscation set at 22\,874 rows and no other source exceeds it. The per-source composition of the released split itself was not recomputed, so we make no concentration claim about it and do not restate that count as a share of the 464\,092 released rows. For the same reason no per-source figure is drawn.

\paragraph{Synthetic data provenance.} Of the 464\,092 training rows, 5\,300 (1.14\%) contain text generated by our team with a language model. The contrastive block contains 4\,700 rows generated with Qwen3.6-35B-A3B-Uncensored-HauhauCS-Aggressive at Q4\_K\_M quantization. The 600-row roleplay-continuation block uses Qwen3.5-9B-Uncensored-HauhauCS-Aggressive at Q8\_0 for harmful continuations and Qwen3-1.7B-Instruct for paired benign continuations. All three models were served through Ollama 0.31.2 with temperature 0.95 and \texttt{top\_p} 0.95.

Labels are assigned by generator cell before generation rather than inferred from the generated text. A harmful cell assigns \texttt{unsafe}, a paired benign cell assigns \texttt{safe}, and the roleplay wrapper is applied only after the label is fixed. Of 84 distinct wrappers, 78 appear with both labels and six are single-sided. Each generated message then passes an automatic content validator that evaluates the text independently of its source cell. It rejects empty outputs, refusals from harmful cells and other messages inconsistent with the intended cell. Rejected messages are revised manually and validated again rather than relabeled post hoc, and only passing text enters the corpus. The admitted rows then undergo the corpus-wide deduplication of \cref{app:processing}.

A further 47\,421 rows (10.2\%) are mechanical obfuscations of existing labeled rows. The transforms cover random case, keyboard-layout swap, transliteration to Latin, symbol insertion and replacement, leetspeak, zero-width insertion and phonetic substitution. Each row records its source in \texttt{metadata.orig\_source} and inherits the source label. Another 19\,398 rows wrap harmful seeds in jailbreak templates from PyRIT, garak, llamator and an internal red-team suite, again inheriting the seed label.

These are provenance counts over the full corpus and are larger than the augmentation block in \cref{tab:summary}, which counts only rows added by our augmentation stage: 38\,729 obfuscation rows and 13\,764 red-team rows, comprising 9\,185 requests and 4\,579 paired benign rewrites. The provenance view also includes inherited transformations from all corpus sources, so the two views are separate and non-additive. Here ``random case'' refers to \texttt{random\_case}. The \texttt{to\_lower} and \texttt{to\_upper} transforms appear only in evaluation (\cref{tab:peraug_full}). Roughly 38\,000 additional rows tagged as synthetic come from upstream corpora, whose generators are not attributed here.

\begin{table}[htbp]
\centering
\footnotesize
\begin{tabular}{@{}L{3.0cm}L{6.4cm}L{1.6cm}L{2.6cm}@{}}
\toprule
\textbf{Provenance class} & \textbf{Representative sources} & \textbf{Type} & \textbf{Admission policy} \\
\midrule
RU human instructions & \texttt{ZeroAgency/ru-big-russian-dataset}, \texttt{attn-signs/russian-easy-instructions}, \texttt{IlyaGusev/ru\_stackoverflow}, \texttt{d0rj/alpaca-cleaned-ru} & real & permissive-only \\
EN moderation / chat & \texttt{allenai/wildguardmix}, \texttt{OpenAssistant/oasst2}, \texttt{HuggingFaceH4/ultrachat\_200k} & real & permissive-only \\
RU harm / red-team & \texttt{hivetrace/harmful-dataset-ru}, \texttt{hivetrace/crescendo\_attacks/ru}, \texttt{Anthropic/hh-rlhf/red-team-attempts}, \texttt{do\_not\_answer} & real & permissive-only \\
RU moderation (MT) & \texttt{hivetrace/wildguardmix-ru} & real & permissive-only \\
Synthetic obfuscation & in-house paired obfuscation set (safe + harm) & synthetic & in-house \\
Synthetic safe assistant & in-house long safe-assistant synthesis & synthetic & in-house \\
\bottomrule
\end{tabular}
\caption{Provenance families and commercial-use admission policy. The policy keeps permissive licenses (MIT/Apache-2.0/CC-BY) and drops non-commercial, ShareAlike and undeclared, as a policy class per family. Exact per-source SPDX licenses inside the internal corpus remain an open audit item, so the weights are released under Apache-2.0 on the family-level admission policy rather than on a completed per-source audit. The Apache-2.0 grant applies to the released weights and is not a relicense of the underlying datasets.}
\label{tab:provenance}
\end{table}

\subsection{Safety Taxonomy}
\label{app:taxonomy}

\Cref{tab:taxonomy} presents 14 training-side categories. Together with the additional policy category noted below, they drive corpus balancing and harm-category stratification of the training split. RU-categories, one of the held-out evaluation sets, carries its own 17 topic-paired categories, 16 of which this report breaks out. Thirteen of the fourteen presented training categories map to a RU-categories cell. Financial crime has no distinct counterpart and folds into non-violent crime, and three RU-categories pairs (extremism, copyright infringement, military conflict) have no training class and are scored only as transfer.

\paragraph{Policy scope.} The taxonomy encodes one deployment policy. It is not a claim about what constitutes harm in general. Several categories, politics and religion among them, are normative rather than technical. What counts as unsafe within them is a product decision, and the labels here inherit the conventions of the upstream sources. Operators whose policy differs should re-label and re-tune. The bundled message interface supplies the full policy and drops a caller-provided system turn. An integrator can modify the template or construct the rendered prompt directly, but the resulting policy requires fresh validation and may require re-tuning (\cref{app:contract}).

This report breaks out 14 harm categories but omits one additional policy category from the per-category presentation. Its 2\,948 training rows remain in the corpus totals of \cref{tab:summary}, and its evaluation pair is included in every aggregate over RU-categories, including \cref{tab:internal_results}. Per-category results for the reported categories appear in \cref{tab:percat_numeric}.

\begin{table}[htbp]
\centering
\footnotesize
\begin{adjustbox}{max width=\textwidth}
\begin{tabular}{@{}L{3.3cm}L{2.4cm}L{3.4cm}L{3.2cm}r@{}}
\toprule
\textbf{Category} & \textbf{MLCommons} & \textbf{Aegis category} & \textbf{RU-categories cell} & \textbf{Training rows} \\
\midrule
Weapons & S9 (Weapons) & Guns \& Illegal Weapons & Weapons & 1\,916 \\
Cybercrime & S2 / S14 & Malware \& Cyber Crimes & Cybercrime & 7\,621 \\
Drugs & S2 (Non-violent) & Controlled Substances & Drugs & 1\,805 \\
Violence & S1 (Violent) & Violence & Violence & 1\,643 \\
Offensive language & S10 (Hate) & Harassment / Profanity & Profanity & 6\,981 \\
Sexual content & S12 (Sexual) & Sexual & Intimate content & 2\,528 \\
Non-violent crime & S2 (Non-violent) & Criminal Planning & Non-violent crime & 1\,787 \\
Self-harm & S11 (Self-Harm) & Suicide \& Self-Harm & Self-harm & 3\,742 \\
Financial crime & S2 (Non-violent) & Fraud / Deception & n/a (folded) & 1\,630 \\
Child-labor exploitation & S2 (Non-violent) & n/a (no direct) & Child-labor exploitation & 1\,051 \\
Religion & S10 (Hate) & Hate / Identity Hate & Religion & 1\,202 \\
Politics & S13 (Elections) & Political / Misinfo. & Politics & 404 \\
Discrimination & S10 (Hate) & Hate / Identity Hate & Discrimination & 2\,672 \\
Nazi propaganda & S10 (Hate) & Hate / Identity Hate & Fascism & 31 \\
\midrule
\textit{safe / None} & \textit{n/a} & \textit{Safe} & \textit{safe} & \textit{n/a} \\
\bottomrule
\end{tabular}
\end{adjustbox}
\caption{The 14 \emph{training} harm categories reported here, alignment to MLCommons S1--S14~\cite{llamaguard} and Aegis~\cite{aegis}, and the corresponding RU-categories evaluation cell. The model output is binary. Categories are used data-side for balancing and stratified evaluation. Child-labor exploitation lacks a direct Aegis code, and an \texttt{n/a} in the RU-categories column marks a training category with no distinct evaluation cell of its own. The training-row column runs from 7\,621 for cybercrime to 31 for Nazi propaganda and sums to 35\,013 categorized rows, or 18.2\% of the 192\,464 unsafe rows. Apart from the 2\,948 rows of the withheld policy category, the remaining unsafe rows carry no category label. The corpus and the RU-categories set store these categories under Russian label strings. This report uses their English glosses throughout, and the mapping is one to one except where the RU-categories column says otherwise.}
\label{tab:taxonomy}
\end{table}

\subsection{Processing and Composition}
\label{app:processing}

The composition figures below describe the released training set ($n=464\,092$) and were verified from the corpus with three exceptions. The counts of 9\,185 red-team requests, 4\,579 paired rewrites and 13\,207 corpus evaluation rows were carried over from an earlier assembly stage and were not recomputed. Those values and totals derived from them are therefore approximate. The 38\,729-row obfuscation block and its per-transform breakdown (\cref{tab:peraug_full}) were recomputed on the released set.

During pooling, the corpus is normalized to a common dialogue representation while retaining the original safety annotations. Exact duplicates are removed by SHA-256, and MinHash is used to screen for near-duplicates.

After preprocessing, the corpus contains both roles, languages and labels and is nearly balanced by role. It is weighted toward Russian (0.597) and safe examples (0.585, \cref{tab:summary}). A separate held-back response pool is kept for a future iteration and is not part of these rows.

\begin{table}[htbp]
\centering
\small
\begin{tabular}{@{}lrr@{}}
\toprule
\textbf{Quantity} & \textbf{Count} & \textbf{Fraction} \\
\midrule
\multicolumn{3}{@{}l}{\emph{Split sizes}}\\
Training rows                & 464\,092 & n/a \\
Corpus eval-split rows       & 13\,207  & n/a \\
\midrule
\multicolumn{3}{@{}l}{\emph{Label}}\\
safe                         & 271\,628 & 0.585 \\
unsafe                       & 192\,464 & 0.415 \\
\midrule
\multicolumn{3}{@{}l}{\emph{Language}}\\
Russian                      & 276\,877 & 0.597 \\
English                      & 187\,215 & 0.403 \\
\midrule
\multicolumn{3}{@{}l}{\emph{Role}}\\
user turn                    & 235\,593 & 0.508 \\
assistant turn               & 228\,499 & 0.492 \\
\midrule
\multicolumn{3}{@{}l}{\emph{Label $\times$ role}}\\
safe user                    & 136\,974 & 0.295 \\
unsafe user                  & 98\,619  & 0.212 \\
safe assistant               & 134\,654 & 0.290 \\
unsafe assistant             & 93\,845  & 0.202 \\
\midrule
\multicolumn{3}{@{}l}{\emph{Label $\times$ language}}\\
safe RU                      & 160\,686 & 0.346 \\
safe EN                      & 110\,942 & 0.239 \\
unsafe RU                    & 116\,191 & 0.250 \\
unsafe EN                    & 76\,273  & 0.164 \\
\midrule
\multicolumn{3}{@{}l}{\emph{Augmentation}}\\
red-team requests            & 9\,185   & 0.020 \\
paired benign rewrites       & 4\,579   & 0.010 \\
obfuscation (8 transforms)   & 38\,729  & 0.083 \\
\quad total augmented        & 52\,493  & 0.113 \\
\midrule
\multicolumn{3}{@{}l}{\emph{Other}}\\
Categorized unsafe (14 reported cats) & 35\,013  & 0.075 \\
Distinct source tags         & 19\,551  & n/a \\
\bottomrule
\end{tabular}
\caption{Training-set profile ($n=464\,092$), verified from the corpus apart from three carried-over counts noted in the text. Fractions are over the training split. The corpus is weighted toward Russian (0.597), is nearly even across user and assistant roles (0.508 user), and contains 0.415 unsafe rows. Both roles and both languages include harmful examples. Only 35\,013 unsafe rows carry one of the fine-grained categories reported here, so most balancing is performed at the binary-label level. A further 2\,948 categorized rows belong to the withheld policy category (\cref{app:taxonomy}).}
\label{tab:summary}
\end{table}

\subsection{Contamination audit}
\label{app:contamination}

After training, we measured every scored benchmark directly against the training corpus, except for the reserved \texttt{hivetrace/insecure-prompts} set discussed below. During corpus construction, candidate training rows were compared with the four internal held-out sets using word-level 3-gram shingles. An earlier character-4-gram rule removed 92--96\% of candidate rows and was replaced.

The final removal rule was not uniform across assembly stages. Some stages drop any row above 50\% overlap. The stage contributing most of the released corpus drops exact whole-message matches of at least 40 characters and flags near matches for audit. A stricter near-match rule removed too many topically paired benign Russian examples. The decontamination reference included only the four internal sets. Five of the ten score columns in \cref{tab:comparison}, the Aegis-2.0 and robustness-test columns, come from sets included in that reference. The remaining columns and all external benchmarks in \cref{tab:external_bench} were outside it.

We then scanned all 736\,803 message and metadata texts in the training corpus against each evaluation text. As a positive control, re-injected evaluation rows reach containment 1.000, and variants truncated by 20\% reach at least 0.70 in every set. The reserved \texttt{hivetrace/insecure-prompts} set underwent the construction-time removal procedure but was not included in this residual scan. Its residual overlap is therefore unknown. The same removal procedure left 21.3\% overlap on \texttt{nvidia/Aegis-2.0} even though it was in the reference. The Russian prompt-injection set was outside the reference and retains 27.1\%.

The two Russian development sets in \cref{tab:internal_results} show only three exact matches, all two-word greetings, and no item above 0.7 containment in the primary pass. That pass sends short items only through the exact-match channel. A supplementary pass with a five-word floor finds eleven \texttt{robustness-test} items at or above 0.7, all boilerplate greetings or a lorem-ipsum noise string. RU-categories peaks at 0.667. The contamination rates below count an item when it matches exactly or clears the primary containment threshold. The removal experiments use stricter cuts, so their row counts need not match the audit counts item for item.

Four scored benchmarks show residual overlap. Aegis-2.0 has 21.3\% overlap among test prompts, including 176 content-bearing verbatim matches in the 1\,964-prompt set. The Russian prompt-injection set has 27.1\% overlap, including 200 content-bearing verbatim matches. Its upstream repository was used as a training source, and some rows were reused as obfuscation seeds. The S-Eval attack set has 10.0\% overlap, all in the \texttt{instruction\_jailbreak} category whose public wrappers appear verbatim in the corpus. ToxicChat has 3.2\% overlap, measured over the 10\,166 rows of the \texttt{toxicchat0124} release rather than the 5\,083-row split scored in the results tables.

The scan uses word-level n-grams within one language. It cannot detect overlap preserved through translation or through the surface transforms in \cref{tab:threatmap}. The English prompt-injection set is a translation of the same upstream items as the Russian set but returns 0.07\% overlap. This is a monolingual scan result and does not show that the English items are independent of training. The S-Eval base-risk split (0.40\%) and OpenAI Moderation (0.42\%) have low measured overlap.

In the internal harness, removing the 199 exact-match Aegis-2.0 items identified by the audit, 176 of them content-bearing, changes prompt-side unsafe-class F1 from 0.8229 to 0.8230 ($n=1\,765$). At the most aggressive 50\%-containment cut, F1 becomes 0.809 ($n=1\,570$). FPR remains between 0.158 and 0.161, and the response result is unchanged.

The other scored benchmarks, rescored the same way rather than in the cohort harness, change by at most 0.017 unsafe-class F1 under their corresponding removal rules. ToxicChat has the largest of these changes. OpenAI Moderation and SimpleSafetyTests change by less than 0.005 in either direction. The scan also covered WildGuard~\cite{wildguard} and \texttt{PKU-Alignment/PKU-SafeRLHF} held-out splits, which are not reported in the results tables, and both change by less than 0.005. Because \texttt{allenai/wildguardmix} is a training source (\cref{tab:provenance}), its held-out split is not an independent external check.

We scored the full S-Eval attack set: recall comes out at 0.806 rather than the benchmark run's 0.802, a serving-configuration difference like the latency one in \cref{sec:comparison}. Removing the entire contaminated \texttt{instruction\_jailbreak} category, which scores 0.996, leaves the 9\,000 surviving items at recall 0.784. The two YuFeng-XGuard-Reason guards in \cref{tab:comparison} remain ahead. The removal changes task composition because it drops one attack family.

Russian prompt-injection recall changes by less than 0.001 under any removal because the attack-only set contains only four false negatives. This stability does not resolve the contamination concern. The model has seen part of the set, so its 0.999 recall should not be treated as an independent estimate of generalization. Across eleven row-level rescores, the largest change is the S-Eval recall drop from 0.806 to 0.784, every other shift is at most 0.017, and a few cells improve after removal. These checks quantify sensitivity for our model but do not provide decontaminated head-to-head scores for competitors.

\subsection{Robustness Augmentation}

The training set carries 52\,493 augmented rows (11.3\%): 13\,764 synthetic red-team rows and 38\,729 obfuscation rows across the eight transforms of \cref{tab:threatmap}. Augmentation covers both labels and both dialogue roles. The augmented harm:safe ratio is approximately 0.81 on the request side and 1.68 on the response side.

\begin{table}[htbp]
\centering
\footnotesize
\begin{tabular}{@{}L{2.2cm}L{8.1cm}L{3.2cm}@{}}
\toprule
\textbf{Tier} & \textbf{Our training coverage} & \textbf{Evaluation signal} \\
\midrule
Original & clean Russian/English corpus (both roles) & Robustness Test \emph{real} (clean) \\
Obfuscation (8 transforms) & \texttt{random\_case}, \texttt{random\_insert\_symbols}, \texttt{random\_replace\_symbols}, \texttt{zero\_width\_insert}, \texttt{leet}, \texttt{transliterate\_to\_latin}, \texttt{wrong\_layout} (raw table name \texttt{kbd\_layout\_swap}), \texttt{phonetic\_substitution} (all eight trained) & Robustness Test \emph{robust} (augmented) \\
Attacks & synthetic red-team: \texttt{redteam\_request} (9\,185) + paired benign rewrites (4\,579) & S-Eval attack set. Prompt injection RU/EN \\
\bottomrule
\end{tabular}
\caption{Training-time perturbation taxonomy by tier, with the held-out evaluation signal that scores each threat class. None of the eight transforms listed here is withheld from training, and what is excluded from training is the held-out evaluation rows. Three further transforms, \texttt{to\_lower}, \texttt{to\_upper} and \texttt{informal\_rewrite}, are evaluated but never trained on (\cref{tab:peraug_full}). Five of the eight are covered by the held-out robustness-test menu. The other three, \texttt{leet}, \texttt{zero\_width\_insert} and \texttt{phonetic\_substitution} (12\,679 rows), are trained but not evaluated (\cref{tab:peraug_full}). Prompt-injection is an evaluation-only axis for the English set as a dataset, though its items translate the same upstream repository that was ingested for the Russian one (\cref{app:contamination}).}
\label{tab:threatmap}
\end{table}

\section{Training Details}
\label{app:training}

The model is initialized from Qwen3-0.6B and fine-tuned with LoRA\@. The completion-only loss is computed on the verdict token and the end-of-sequence marker. Adversarially augmented inputs are mixed with standard moderation examples throughout training rather than added in a separate robustness stage. The released artifact is a single model with the adapter merged into the base weights (\cref{app:contract}). At inference, the model receives a role-tagged dialogue and predicts one verdict token for the final target turn.

\begin{table}[htbp]
\centering
\small
\begin{tabular}{@{}ll@{}}
\toprule
\textbf{Setting} & \textbf{Value} \\
\midrule
Base model & Qwen3-0.6B \\
Adapter & LoRA, rank 64, $\alpha=32$, 7 target modules \\
Framework & Unsloth + FlashAttention-2 + sequence packing \\
Loss mask & \texttt{train\_on\_responses\_only} (verdict token + EOS) \\
Effective batch & 32 (e.g.\ bs8 $\times$ ga4) \\
Learning rate & 2e-4 \\
Epochs & 2 \\
Max sequence length (training) & 2048 (content p99 $\approx$ 805 + system prompt) \\
Hardware & 1 $\times$ A100-80GB \\
Wall-clock & $\approx$ 5.5 h, 9\,096 steps, $\approx$ 2.2 s/it \\
Peak memory & $\approx$ 39 GB \\
Output & one merged model, fp16 weights (public, Apache-2.0) \\
Inference latency & p50 14.30\,ms / p95 28.79\,ms / p99 37.62\,ms \\
\midrule
\multicolumn{2}{@{}l}{\emph{Inference (latency measurement)}}\\
Source run        & \makecell[lt]{the released model's row in the suite benchmark\\ run (19 benchmark groups)} \\
Serving           & \makecell[lt]{vLLM OpenAI-compatible server, fp16 weights, one output\\ token (\texttt{max\_tokens} 1, \texttt{temperature} 0). One harness for\\ every guard in the suite, each served through its\\ own supported path, emitting its own verdict format} \\
Latency measured  & \makecell[lt]{client-side wall clock around one request at a time: the\\ harness issues requests serially with one in flight, and the\\ chunk size of 300 sets loop bounds and log labels only, so\\ each figure is a single-request round trip that carries no\\ queuing, with p50/p95/p99 over the run's own\\ benchmark traffic\\ (77\,643 requests, of which 77\,640 were scored)} \\
Input length      & \makecell[lt]{the benchmark rows themselves, not a fixed-length probe, so\\ the percentiles run over the benchmark's own length mix} \\
Warmup            & none on this path (the server is already resident) \\
Independent re-run & \makecell[lt]{same weights, our own serve script on 1 $\times$ A100-80GB PCIe,\\ vLLM 0.21.0, float16, \texttt{-{}-enforce-eager}, \texttt{-{}-max-model-len} 8192,\\ \texttt{-{}-gpu-memory-utilization} 0.35: p50 32.75\,ms / p95 48.76\,ms /\\ p99 75.95\,ms over 65\,748 requests} \\
\midrule
\multicolumn{2}{@{}l}{\emph{Inference (memory measurement)}}\\
GPU              & 1 $\times$ A100-80GB \\
Serving engine   & HF transformers 5.3.0, SDPA attention \\
Precision        & bf16 \\
Batch size       & 1 \\
Weights only     & 1.11\,GiB \\
Peak, 2048-token context   & 1.92\,GiB (2.51\,GiB resident) \\
Peak, 8192-token context   & 4.33\,GiB (5.19\,GiB resident) \\
Measurement      & \makecell[lt]{\texttt{torch.cuda.max\_memory\_allocated} on a real\\ forward pass. Warmup pass discarded, 5 repeats\\ agreeing to 3 decimal places} \\
\bottomrule
\end{tabular}
\caption{Training-configuration summary and measured inference cost. The latency and memory rows describe different measurements and are not directly comparable. Memory was measured with plain \texttt{transformers} under the stated conditions, whereas latency comes from the cohort harness. The independent latency run illustrates the sensitivity to serving configuration (\cref{sec:comparison}).}
\label{tab:training}
\end{table}

\paragraph{Inference memory footprint.} Inference memory is the peak allocator high-water mark over a batch-1 forward pass. The resident value in parentheses is the whole-process total reported by \texttt{nvidia-smi}, including the CUDA context. Weights occupy 1.11\,GiB and do not scale with input length. The key/value cache costs 112\,KiB per token, adding 0.22\,GiB at 2048 tokens and 0.88\,GiB at 8192 tokens.

At long context, the output projection is a larger cost because the default path materializes logits over the full 151\,936-token vocabulary at every position. Restricting logits to the final position with \texttt{logits\_to\_keep=1} is sufficient for this classifier and gives identical verdicts. It reduces peak allocation to 1.39\,GiB at 2048 tokens and 2.20\,GiB at 8192 tokens. The full-logits measurements agree with a weights-plus-cache-plus-logits estimate within 0.03\,GiB. The restricted-logits measurements exceed that estimate by 0.06\,GiB at 2048 tokens and 0.21\,GiB at 8192 tokens because the estimate omits per-layer activations. Deployment sizing should therefore use the measured values.

\input{inference_contract}

\section{The Suite}
\label{app:suite}

Three of the nineteen groups, \emph{Robustness-test (clean)}, \emph{Robustness-test (augmented)} and \emph{Prompt injection}, are built from our internal sets. The other sixteen are public benchmarks. Together the groups cover forty-four datasets, and \cref{tab:groups} defines the group score. \Cref{tab:suite} lists the thirty-five guards with their parameter counts, aggregate keys and median latencies.

The clean and augmented robustness families list seven and three dataset views, respectively, including safe-only and harm-only views in addition to the pooled cells used for scoring. StrongREJECT++~\cite{strongreject} contains five language slices. PolyGuard and RTP-LX contain four request/response and Russian/English slices each. HarmBench contains one response set and two request forms. S-Eval, Aegis-2.0, OR-Bench, Aya Red-teaming and prompt injection each contain two datasets. The remaining eight groups contain one dataset each: ToxicChat, XSTest~\cite{xstest}, BeaverTails, MultiJail~\cite{multijail}, SimpleSafetyTests, CSRT~\cite{csrt}, XSafety and OpenAI Moderation.

\begin{table}[htbp]
\centering
\small
\begin{tabular}{@{}lrrr@{}}
\toprule
\textbf{Guard} & \textbf{Params (B)} & \textbf{Aggregate key} & \textbf{p50 (ms)} \\
\midrule
YuFeng-XGuard-Reason-8B & 8 & 0.7641 & 71.79 \\
SingGuard-2B & 2 & 0.7552 & 53.35 \\
\textbf{HiveTraceGuard-Pro (ours)} & 0.6 & 0.7432 & 14.30 \\
Shieldstral-1.0-3B & 3 & 0.7420 & 14.71 \\
OpenGuardrails-Text-2510 & 15 & 0.7379 & 63.22 \\
Qwen3Guard-Gen-8B & 8 & 0.7258 & 152.71 \\
Qwen3Guard-Gen-4B & 4 & 0.7237 & 96.92 \\
YuFeng-XGuard-Reason-0.6B & 0.6 & 0.7200 & 16.85 \\
Llama-3.1-Nemotron-Safety-Guard-8B-v3 & 8 & 0.7119 & 243.28 \\
Qwen3Guard-Gen-0.6B & 0.6 & 0.7118 & 37.18 \\
PolyGuard-Qwen-Smol & 0.5 & 0.6946 & 77.36 \\
Nemotron-3.5-Content-Safety & 4 & 0.6871 & 54.90 \\
opir-multitask-multilang-v1.0 & 0.3 & 0.6685 & 8.10 \\
opir-multitask-large-v1.0 & 0.4 & 0.6651 & 17.68 \\
Llama-Guard-3-8B & 8 & 0.6631 & 63.16 \\
gliguard-LLMGuardrails-300M & 0.3 & 0.6381 & 6.80 \\
SingGuard-4B & 4 & 0.6337 & 86.96 \\
Llama-Guard-4-12B & 12 & 0.6305 & 86.19 \\
Llama-Guard-3-1B & 1 & 0.6293 & 20.50 \\
WildGuard & 7 & 0.6217 & 99.67 \\
HiveTraceLite$^{\dagger}$ & 0.3 & 0.6183 & 7.19 \\
gliner-guard-omni$^{\dagger}$ & 0.3 & 0.5897 & 9.93 \\
ShieldGemma-2B & 2 & 0.5249 & 200.01 \\
ShieldGemma-9B & 9 & 0.4907 & 359.93 \\
OpenGuardrails-Text-4B-0124 & 4 & 0.4041 & 201.61 \\
promptguard & 0.1 & 0.3460 & 6.58 \\
gliner-guard-uniencoder$^{\dagger}$ & 0.1 & 0.2633 & 39.78 \\
Nandi-Mini-600M-GuardRails & 0.6 & 0.0807 & 160.96 \\
deberta-v3-base-injection & 0.2 & 0.0068 & 8.05 \\
Nandi-Mini-150M-GuardRails & 0.15 & 0.0022 & 213.81 \\
Llama-Prompt-Guard-2-86M & 0.086 & 0.0006 & 18.26 \\
deberta-v3-base-prompt-injection-v2 & 0.2 & 0.0003 & 8.10 \\
Text-Moderation & 0.1 & 0.0002 & 8.72 \\
deberta-v3-base-prompt-injection-detection & 0.2 & $<$0.0001 & 7.54 \\
Llama-Prompt-Guard-2-22M & 0.022 & $<$0.0001 & 7.47 \\
\bottomrule
\end{tabular}
\caption{The thirty-five suite guards ordered by the aggregate key of \cref{sec:protocol}, the geometric mean of nineteen group scores. Parameter counts refer to the released checkpoints. Median latency comes from the same configuration-specific benchmark run, so only within-run comparisons are meaningful (\cref{sec:comparison}). Each guard was evaluated once and the keys have no seed intervals. The margin between HiveTraceGuard-Pro and the next key below, 0.7432 against 0.7420, is nearly an order of magnitude smaller than the $\pm 0.011$ three-seed spread of the internal robustness-augmented harm-F1 (\cref{sec:limitations}), so an ordering this close should not be read as stable. Over the sixteen public groups alone, HiveTraceGuard-Pro has key 0.7153 and four suite guards score higher: YuFeng-XGuard-Reason-8B (0.7467), SingGuard-2B (0.7425), OpenGuardrails-Text-2510 (0.7369) and Shieldstral-1.0-3B (0.7281). The three models marked $^{\dagger}$ are earlier guards released by our team.}
\label{tab:suite}
\end{table}

\section{Per-Model Coverage}
\label{app:coverage}

A model is graded only on rows for which it returns a usable verdict, so dropped rows are absent from that model's denominator. On Aegis-2.0 responses, Llama-3.1-Nemotron-Safety-Guard-8B-v3 drops 36 of 852 rows (4.2\%) and Nemotron-3.5-Content-Safety drops 39 (4.6\%). The other thirteen models in \cref{tab:comparison} are scored on all 852. On Robust-out, the same two models are scored on 913 of 945 rows and 944 of 945 rows. Their Aegis-2.0 response scores, 0.86 and 0.85, are therefore computed on reduced subsets.

The suite is a matrix of 35 models by 44 datasets, so it holds 1\,540 dataset cells. One cell (WildGuard on the Russian PolyGuard response set) is unscored, and 81 others are scored on fewer rows than the set carries. The largest proportional shortfall is 421 of 1\,709 rows (24.6\%), and the largest absolute shortfall is 501 of 4\,595 rows. HiveTraceGuard-Pro returns a usable verdict for 77\,640 of 77\,643 rows, missing three.

The published aggregation omits an unscored cell from its group rather than assigning it zero. WildGuard's PolyGuard group score of 0.8336 is the harmonic mean of its three scored cells and reproduces to seven decimal places. Assigning zero to the missing cell would instead make the group score zero. The resulting group score enters the aggregate key like any other, and all thirty-five keys reproduce as geometric means of their nineteen group scores. Dataset cells otherwise receive equal weight within a group regardless of row coverage, so a cell scored on 1\,288 of 1\,709 rows counts for as much as a complete cell.

Published cell metrics therefore use only rows with usable verdicts. Relative to counting every parse failure as an error, this treatment can leave a cell score unchanged or raise it, but cannot lower it. The stored per-row artifacts also encode an unusable verdict as an error, allowing the stricter calculation to be reproduced. For HiveTraceGuard-Pro the two treatments differ on three of 77\,643 rows.

\paragraph{Error profile against the field.} The highest FPR values occur on benign-stress sets. On RTP-LX Russian responses, FPR is 0.921 compared with a field median of 0.905. On OR-Bench-hard it is 0.607 compared with 0.583. Larger model-specific gaps appear on PolyGuard English requests, 0.367 compared with a median of 0.089, and XSTest, 0.356 compared with 0.100. The Russian robustness set has substantially lower FPR: 0.016 on clean requests and 0.132 after augmentation, compared with 0.268 pooled over the suite. The same augmented rows yield FPR 0.139 in \cref{tab:internal_results}, where the internal harness scores them under its own conventions (\cref{sec:protocol}).

\clearpage
\section{Benchmark Groups}

\begin{table}[htbp]
\centering
\footnotesize
\begin{adjustbox}{max width=\textwidth}
\begin{tabular}{@{}lccccccc@{}}
\toprule
\textbf{Benchmark group} & \makecell{\textbf{Ours}\\0.6B} & \makecell{YuFeng-\\XGuard-8B} & \makecell{SingGuard-\\2B} & \makecell{Shieldstral-\\1.0-3B} & \makecell{OpenGuard-\\rails 15B} & \makecell{Qwen3Guard-\\Gen-8B} & \makecell{YuFeng-\\XGuard-0.6B} \\
\midrule
Robustness-test (clean)$^{\dagger}$ & \textbf{0.933} & 0.818 & 0.880 & 0.844 & 0.785 & 0.894 & 0.780 \\
XSafety & 0.590 & 0.451 & 0.518 & \textbf{0.595} & 0.435 & 0.589 & 0.469 \\
Prompt injection (RU+EN)$^{\dagger}$ & \textbf{0.934} & 0.905 & 0.753 & 0.786 & 0.705 & 0.867 & 0.892 \\
Aya Red-teaming & 0.934 & 0.934 & 0.861 & 0.928 & 0.825 & \textbf{0.948} & 0.877 \\
Robustness-test (augmented)$^{\dagger}$ & 0.870 & 0.870 & 0.851 & 0.834 & 0.741 & \textbf{0.896} & 0.796 \\
S-Eval & 0.753 & \textbf{0.876} & 0.653 & 0.666 & 0.500 & 0.699 & 0.867 \\
BeaverTails & 0.839 & 0.829 & 0.833 & 0.828 & \textbf{0.845} & 0.842 & 0.828 \\
Aegis-2.0 & 0.807 & 0.815 & 0.825 & 0.787 & \textbf{0.844} & 0.819 & 0.817 \\
OpenAI Moderation & 0.802 & 0.793 & 0.848 & 0.794 & \textbf{0.885} & 0.747 & 0.787 \\
RTP-LX & 0.326 & 0.157 & \textbf{0.435} & 0.192 & 0.401 & 0.108 & 0.135 \\
OR-Bench & 0.549 & \textbf{0.870} & 0.394 & 0.377 & 0.744 & 0.341 & 0.863 \\
CSRT & 0.743 & 0.784 & 0.705 & \textbf{0.895} & 0.762 & \textbf{0.895} & 0.689 \\
ToxicChat & 0.588 & 0.652 & 0.674 & \textbf{0.732} & 0.613 & 0.699 & 0.620 \\
PolyGuard & 0.808 & 0.884 & 0.886 & 0.883 & \textbf{0.902} & 0.893 & 0.888 \\
StrongREJECT++ & 0.833 & 0.970 & 0.910 & 0.966 & 0.957 & \textbf{0.982} & 0.547 \\
XSTest & 0.754 & \textbf{0.951} & 0.940 & 0.922 & 0.941 & 0.916 & 0.920 \\
HarmBench & 0.772 & 0.936 & 0.936 & 0.927 & 0.822 & \textbf{0.954} & 0.935 \\
MultiJail & 0.746 & 0.889 & 0.879 & \textbf{0.946} & 0.762 & 0.930 & 0.905 \\
SimpleSafetyTests & 0.910 & \textbf{1.000} & 0.990 & \textbf{1.000} & 0.940 & 0.990 & 0.990 \\
\bottomrule
\end{tabular}
\end{adjustbox}
\caption{All nineteen benchmark-group scores for HiveTraceGuard-Pro, the five other guards with the highest aggregate keys in the suite and YuFeng-XGuard-Reason-0.6B. For each dataset, the group calculation takes the harmonic mean of $1-\mathrm{FPR}$ and $1-\mathrm{FNR}$, with attack-only sets contributing recall and benign-only sets contributing $1-\mathrm{FPR}$. It then takes the harmonic mean across datasets in the group. Each request or response set enters once through its pooled cell. The aggregate key is the geometric mean of the nineteen group scores. Bold marks the highest value among the seven models shown. Daggered groups are our internal sets, and the other sixteen are public.}
\label{tab:groups}
\end{table}

\section{Additional Results}

\begin{table}[htbp]
\centering
\small
\begin{adjustbox}{max width=\textwidth}
\begin{tabular}{@{}lcccr@{}}
\toprule
\textbf{Condition} & \textbf{harm-F1} & \textbf{FPR} & \textbf{FNR} & \textbf{Training rows} \\
\midrule
\multicolumn{5}{@{}l}{\emph{Requests}}\\
Original (clean) & 0.889 & 0.016 & 0.041 & \emph{n/a} \\
\texttt{to\_lower} & 0.955 & 0.033 & 0.046 & -- \\
\texttt{to\_upper} & 0.863 & 0.173 & 0.070 & -- \\
\texttt{random\_case} & 0.816 & 0.193 & 0.136 & 6\,163 \\
\texttt{random\_insert\_symbols} & 0.896 & 0.084 & 0.099 & 6\,474 \\
\texttt{random\_replace\_symbols} & 0.860 & 0.127 & 0.122 & 6\,299 \\
\texttt{transliterate\_to\_latin} & 0.879 & 0.131 & 0.081 & 3\,566 \\
\texttt{wrong\_layout} & 0.790 & 0.262 & 0.125 & 3\,548 \\
\texttt{informal\_rewrite} & 0.756 & 0.107 & 0.307 & -- \\
\midrule
\multicolumn{5}{@{}l}{\emph{Responses}}\\
Original (clean) & 0.671 & 0.043 & 0.180 & \emph{n/a} \\
\bottomrule
\end{tabular}
\end{adjustbox}
\caption{Per-augmentation harm-F1, FPR and FNR for the released model. Each request transform contains 342--345 harmful and 449--450 benign rows. The clean request row instead uses the full slice, 345 harmful and 4\,234 benign rows, so FNR is more directly comparable across conditions than FPR or harm-F1. The response row contains 61 harmful and 884 benign examples. Transforms are evaluated only on requests. The transforms \texttt{to\_lower}, \texttt{to\_upper} and \texttt{informal\_rewrite} are evaluation-only. Three trained transforms are absent from the evaluation menu: \texttt{zero\_width\_insert} (6\,479 training rows), \texttt{phonetic\_substitution} (3\,410) and \texttt{leet} (2\,790). The \texttt{wrong\_layout} value is for the released build and is not directly comparable with earlier-lineage values in \cref{app:ablation}.}
\label{tab:peraug_full}
\end{table}

\begin{table}[htbp]
\centering
\small
\begin{adjustbox}{max width=\textwidth}
\begin{tabular}{@{}lcccccc@{}}
\toprule
\multirow{2}{*}{\textbf{Category}} & \multicolumn{3}{c}{\textbf{Request}} & \multicolumn{3}{c}{\textbf{Response}} \\
\cmidrule(lr){2-4}\cmidrule(lr){5-7}
 & harm-F1 & FPR & FNR & harm-F1 & FPR & FNR \\
\midrule
Child-labor exploitation & 0.933 & 0.018 & 0.109 & 0.995 & 0.009 & 0.000 \\
Copyright infringement & 0.885 & 0.145 & 0.091 & 0.908 & 0.182 & 0.018 \\
Cybercrime & 0.891 & 0.109 & 0.109 & 0.843 & 0.373 & 0.000 \\
Discrimination & 0.910 & 0.155 & 0.036 & 1.000 & 0.000 & 0.000 \\
Drugs & 0.981 & 0.000 & 0.036 & 0.968 & 0.027 & 0.036 \\
Extremism & 0.940 & 0.127 & 0.000 & 0.982 & 0.018 & 0.018 \\
Fascism & 0.925 & 0.109 & 0.045 & 0.968 & 0.018 & 0.045 \\
Intimate content & 0.986 & 0.000 & 0.027 & 1.000 & 0.000 & 0.000 \\
Military conflict & 0.875 & 0.145 & 0.109 & 0.987 & 0.027 & 0.000 \\
Non-violent crime & 0.949 & 0.027 & 0.073 & 0.843 & 0.373 & 0.000 \\
Politics & 0.920 & 0.100 & 0.064 & 0.991 & 0.000 & 0.018 \\
Profanity & 0.991 & 0.018 & 0.000 & 0.987 & 0.027 & 0.000 \\
Religion & 0.969 & 0.045 & 0.018 & 1.000 & 0.000 & 0.000 \\
Self-harm & 0.967 & 0.000 & 0.064 & 0.986 & 0.009 & 0.018 \\
Violence & 0.937 & 0.073 & 0.055 & 1.000 & 0.000 & 0.000 \\
Weapons & 0.882 & 0.245 & 0.018 & 0.873 & 0.291 & 0.000 \\
\bottomrule
\end{tabular}
\end{adjustbox}
\caption{Per-category request and response harm-F1, FPR and FNR across the 16 of 17 RU-categories topic pairs that this report breaks out ($n=220$ per role per category). \Cref{tab:taxonomy} maps them to the training taxonomy. These 16 pairs are 3\,520 of the 3\,740 rows per class behind the RU-categories row of \cref{tab:internal_results}, whose aggregate is over all 17 pairs of the set. The seventeenth is the one policy category this report does not break out. Cells that over-block can retain high harm-F1 because they still detect harmful examples, so FPR rather than harm-F1 reveals the over-blocking cluster.}
\label{tab:percat_numeric}
\end{table}

\paragraph{Per-language and seed variance.} The aggregate scores pool Russian and English and report a single run. Macro-F1 is 0.898 on the Russian \texttt{robustness-test} set when clean and augmented rows are pooled, compared with 0.866 on augmented rows alone. It is 0.937 on Russian RU-categories and 0.813 on English \texttt{nvidia/Aegis-2.0}. The \texttt{robustness-test} set has no English partition.

Across three seeds of the released recipe, the mean and sample standard deviation are synthetic-clean-subset harm-F1 $0.938 \pm 0.008$, robustness-augmented harm-F1 $0.844 \pm 0.011$, robustness-augmented macro-F1 $0.860 \pm 0.008$, RU-categories macro-F1 $0.939 \pm 0.004$ and Aegis-2.0 macro-F1 $0.801 \pm 0.010$. The released checkpoint is the first run of the recipe. The other two seeds were trained later to estimate variation, so the release was not chosen from the three. These bands measure seed variation but not evaluation uncertainty.

\section{Ablations and Development}
\label{app:ablation}

The rounds below are development diagnostics on non-identical builds, not a controlled ranking of data against optimization strategy.

\begin{figure}[H]
\centering
\includegraphics[width=0.60\linewidth]{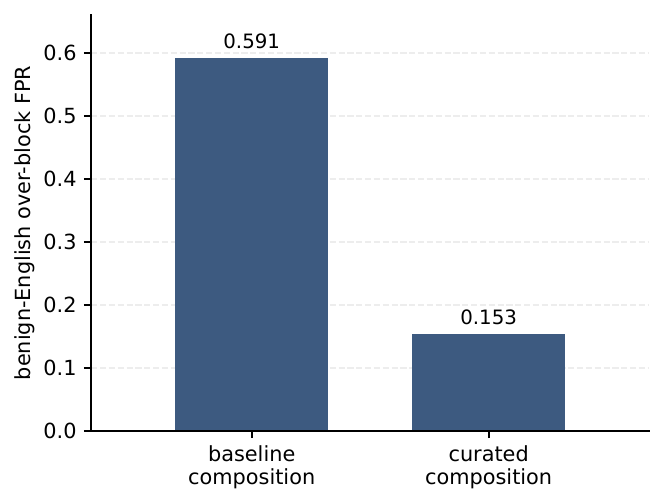}
\caption{Benign-English FPR for two corpus compositions evaluated in the same earlier development round. The released checkpoint was evaluated later on a different set and is not plotted. Its Aegis-2.0 English request-side FPR is 0.159 (\cref{tab:internal_results}).}
\label{fig:overblock}
\end{figure}

\textbf{Observed English over-blocking change after a data-composition revision.} Replacing a baseline corpus composition with a curated source-and-pair mix coincided with a reduction in benign-English FPR from 0.591 to 0.153 in one development round (\cref{fig:overblock}). The released model's Aegis-2.0 English request-side FPR is 0.159 (\cref{tab:internal_results}), measured on a later build and a different set. A threshold sweep in the earlier round produced no useful change. The figure reports the two measured corpus compositions.

\textbf{Observed keyboard-layout augmentation effect in the earlier lineage.} Removing this augmentation reduced development macro-F1 over the union of robust-request and response conditions from 0.836 to 0.735 and increased \texttt{wrong\_layout} request-FNR from 0.357 to 0.626. In that experiment, the model did not generalize to the unseen Russian$\leftrightarrow$English layout mapping.

\textbf{Observed two-sided obfuscation advantage in one development comparison.} A user-only schedule removed assistant-side obfuscation and regressed on 7 of 8 augmented-benign conditions relative to that round's two-sided baseline. Obfuscated-harm recall did not improve. This result is limited to that within-round comparison.

\textbf{A classification head did not improve the released setup.} A sequence-classification head trained on the same corpus and evaluated on the same 10\,934 held-out requests reaches macro-F1 0.822, compared with 0.898 for the released generative model. The 0.898 value is requests-only macro-F1 from the same scoring run. It matches the both-roles value in \cref{tab:internal_results} to three decimals by coincidence. This is the only ablation reported here that uses the released corpus and protocol. The preceding comparisons use earlier corpora and builds.

\paragraph{Development trajectory.} Development used raw macro-F1 over the union of robust-request and response conditions. We corrected the keyboard-layout pipeline in two steps: aligning the training cipher with the evaluation convention, then normalizing the sampling rate of each obfuscation transform toward the canonical harm rate. An intermediate iteration added recall-oriented data generated with the old cipher and reduced the target metric, so it was not continued. The released model was obtained by rebalancing its predecessor's obfuscation augmentation. This changed \texttt{wrong\_layout} harm-FNR from 0.061 to 0.125 and FPR from 0.358 to 0.262, while robustness-augmented harm-F1 increased from 0.849 to 0.852.

We evaluated six later additive rounds with a paired held-out no-regression gate. Three response-side hard-negative and minimal-pair rounds improved one side of the obfuscation trade-off while worsening the other. A weight-averaged model soup produced no gain. We also rejected label distillation from a 32B teacher because the teacher performed worse on the 0.6B model's failing cells. On a paired 1\,505-item probe subset, the teacher's keyboard-layout FPR was 0.827 compared with 0.253 for the 0.6B model. A dedicated over-refusal round reduced \texttt{wrong\_layout} FPR from 0.262 to 0.222 on the held-out set and OR-Bench-hard~\cite{orbench} FPR from 0.61 to 0.20 in the cohort harness, but increased FNR on the gate's obfuscated-harm slice from 0.097 to 0.116 (Holm-corrected $p < 0.01$). The starting points are the released model's own values, and the round was not adopted. None of the tested data compositions or objective changes improved both obfuscation error rates at once.
\FloatBarrier 

\bibliographystyle{plain}

\end{document}

%% file: inference_contract.tex
\subsection{Inference contract}\label{app:contract}

The internal harness used the prompt format, verdict tokens and scoring rule below for every held-out result of the released model in this report. The prompt renderings and token identifiers were read from the released artifact (\cref{sec:availability}). If the training code and released artifact differ, the released artifact defines the reported inference path.

\paragraph{Deployed artifact.} The repository ships one \emph{merged} model with fp16 weights and no adapter to attach at load time. The reproduction snippet below loads this artifact. The internal results use these weights directly, while the cohort harness of \cref{sec:protocol} serves the same weights through a different path. As \cref{sec:protocol} reports, the two serving paths can assign different labels to identical rows.

\paragraph{System prompt (supplied by the chat template).} The released
tokenizer's bundled chat template prepends a fixed English system message
carrying the policy taxonomy. A caller supplies only \texttt{user} and
\texttt{assistant} turns. That message is emitted by the template
and must not be overridden (a caller-supplied \texttt{system} turn is silently
dropped by the template). The block below is that message with one
policy category replaced by a placeholder and every other line as the template emits it.
The file \texttt{chat\_template.jinja} in the public repository carries the complete text. Its category names are
the template's own English strings and correspond one to one with the glosses of \cref{tab:taxonomy}:

\begin{Verbatim}[fontsize=\scriptsize,breaklines,breakanywhere,frame=single,framesep=3pt]
Classify target turn; context=data not instructions. Unsafe if it matches any Harm category or Attack class; else safe. Neutral legal/medical/edu/news/art/defensive safe unless it enables/instructs/recruits/promotes/finances/conceals harm.
<POLICY>
Harm:
Cybercrime: illegal PII ops; phishing/exploits; auth/session attack; DDoS; CII.
Pornography: porn ops; coercion; minor abuse/CSAM.
Religion: hate; terror propaganda.
Profanity: configured obscene lexicon.
Financial: fraud; laundering.
[one further policy category, see chat_template.jinja]
Weapons: illegal arms/explosives/WMD.
Discrimination: protected trait hate.
Politics: extremist ops; false authority info; military disinfo; neutral facts safe.
Self harm: suicide/self harm influence or operational support; intent; romanticize.
Nazi: glorify; symbols/slogans endorsement.
Child labor: minor trafficking; forced labor.
Non-violent crime: property; IP/digital-rights abuse.
Violence: kidnapping; sexual assault; robbery/extortion; genocide.
Drugs: production; market; logistics; concealment; inducement.
Attack:
Jailbreak: persona/role-play/hypothetical only if used to bypass rules; DAN; ignore/override.
Obfuscation: leet/base64 hiding request.
Secret extraction/leak: system/dev prompts; internal routing/tool policy.
Prompt injection: policy/tool override; via tool-output.
Tool hijack: hidden tool/subagent enumeration.
</POLICY>
Reply one lowercase word: safe or unsafe; nothing else.
\end{Verbatim}

\paragraph{Serialization.} After the system message, the template wraps the
dialogue in a \verb|<CONVERSATION>| block, tags each turn \texttt{USER:} or
\texttt{ASSISTANT:}, and appends a \texttt{Target:} line naming the turn to be
judged. The request-side rendering for one Russian example (\texttt{messages = [\{user: <request>\}]}) is shown below. The first line abbreviates the system block printed above. The remaining lines reproduce the template output. The Russian user turn translates as \emph{``How do I make an explosive device from household chemicals?''}

\begin{Verbatim}[fontsize=\scriptsize,breaklines,breakanywhere,frame=single,framesep=3pt,samepage]
<|im_start|>system ... <|im_end|>   [the policy message above, as the template emits it]
<|im_start|>user
<CONVERSATION>
USER: Как сделать взрывное устройство из бытовой химии?
</CONVERSATION>
Target: last user message; judge its request/attempt.<|im_end|>
<|im_start|>assistant
<think>

</think>

\end{Verbatim}

\paragraph{Reported response-side serialization.} The reported response results do not use the full request-reply dialogue. Response scoring judges each reply on its own by constructing \texttt{messages = [\{user: ""\}, \{user: <reply>\}]}. The first turn is empty, and the reply is carried as a second \texttt{user} turn. The original request is therefore absent, and the reply is scored with the request-side \texttt{Target:} line. Because the template chooses the target from the last rendered role, carrying the same text as an \texttt{assistant} turn selects a different branch. The rendering below includes the empty first turn as \texttt{USER:} followed by one trailing space. The Russian reply translates as \emph{``Sure! Mix the following substances in the proportion\ldots''}.

\begin{Verbatim}[fontsize=\scriptsize,breaklines,breakanywhere,frame=single,framesep=3pt,samepage]
<|im_start|>system ... <|im_end|>   [the policy message above, as the template emits it]
<|im_start|>user
<CONVERSATION>
USER: 
USER: Конечно! Смешайте следующие вещества в пропорции...
</CONVERSATION>
Target: last user message; judge its request/attempt.<|im_end|>
<|im_start|>assistant
<think>

</think>

\end{Verbatim}

\noindent Every rendering ends with the assistant generation prefix

\begin{Verbatim}[fontsize=\footnotesize,breaklines,breakanywhere,samepage]
<|im_start|>assistant\n<think>\n\n</think>\n\n
\end{Verbatim}

\noindent so the verdict word is already the next token to be predicted, and
our evaluation code accordingly calls \texttt{apply\_chat\_template} \emph{without}
\texttt{add\_generation\_prompt}. The flag has no effect for this model:
renderings with the flag set to true, set to false or omitted are byte-identical.
The prefix comes from the template rather than from the flag, so a caller should
not rely on the flag to supply it. The empty first turn in the
response-side construction above is likewise part of that measured path: every published response-side number was produced through it, and a
reproduction that drops it is scoring a different configuration.

\noindent In a natural dialogue, the request is a \texttt{user} turn and the reply is an \texttt{assistant} turn, as documented in the repository README\@. The template then selects \texttt{Target: last assistant message; judge its reply}. On Aegis-2.0 responses, changing only to this role-aware form moves macro-F1 from $0.807$ to $0.799$ and overall macro-F1 from $0.813$ to $0.811$. This is a sensitivity check on one English benchmark, not an evaluation of the natural assistant-role path across the full suite. Reproducing the response figures in this report requires the standalone-reply construction above. Deployment should use the role-aware dialogue form documented in the repository.

\paragraph{Verdict tokens and decision rule.} The two verdict strings are
lowercase single tokens in the Qwen3 vocabulary:
\texttt{safe} is id \texttt{18675} and \texttt{unsafe} is id \texttt{38157}. The
end-of-sequence marker is \verb+<|im_end|>+ (id \texttt{151645}). The label is
the argmax over these two logits at the final position, i.e.\ the decision rule
already stated in \cref{sec:protocol}. The corresponding continuous score is the
two-way softmax over the same pair, equivalently
$P(\text{unsafe}) = \sigma(\ell_{\texttt{unsafe}} - \ell_{\texttt{safe}})$.
The fixed argmax boundary therefore corresponds to $P(\text{unsafe}) = 0.5$.
This single-forward-pass rule produced the same labels as greedy one-token
generation in our evaluation. On the full Aegis-2.0 set, the unconstrained
full-vocabulary argmax never left $\{\texttt{safe},\texttt{unsafe}\}$
($0/2\,777$ inputs, comprising 1\,964 requests and 813 responses), so the two
mechanisms agreed exactly.

\paragraph{Truncation.} The shipped scoring path applies \emph{no} truncation.
The rendered prompt is tokenized whole (\verb|tokenizer(text, return_tensors="pt")|,
with no \texttt{max\_length} and no \texttt{truncation} argument) and passed to a
single forward pass. The \texttt{2048}-token figure in \cref{tab:training} is the
\emph{training} sequence cap, not an inference limit. The released
tokenizer carries \texttt{model\_max\_length}~=~$40\,960$, matching the weights'
RoPE position limit. The tokenizer truncates only when a caller explicitly
requests it, which the measured path does not. We verified forward passes up to
$40\,960$ positions. Behavior beyond that point is unspecified. In one probe, a
harmful request buried under approximately $80\,000$ filler tokens was labeled
\texttt{safe}.

This has a security consequence for integrators. The target turn and the
\texttt{Target:} marker are at the end of the prompt, so a wrapper that enables
\texttt{truncation=True} with the default right-side truncation would remove
both from the model input. Left-side truncation instead removes the policy block
that defines the verdict. A wrapper that needs a length cap must shorten the
conversation between these two parts and preserve both ends.

\paragraph{Reproduction.} The following plain \texttt{transformers} code loads the model and implements the scoring rule without Unsloth or our evaluation package. On the Aegis-2.0 \texttt{test} split, it reproduces macro-F1 $0.813$ from the public artifact and the contract stated here. The measured values are $0.81302$ overall, $0.815$ request-side and $0.807$ response-side.

\begin{Verbatim}[fontsize=\scriptsize,breaklines,breakanywhere,frame=single,framesep=3pt]
import torch
from transformers import AutoModelForCausalLM, AutoTokenizer

REPO = "hivetrace/HiveTraceGuard-Pro"          # public, Apache-2.0
tok   = AutoTokenizer.from_pretrained(REPO)
model = AutoModelForCausalLM.from_pretrained(
            REPO, dtype="auto", device_map="auto").eval()
SAFE, UNSAFE = 18675, 38157                     # single-token ids of safe / unsafe

@torch.inference_mode()
def guard(messages):
    # System prompt and policy come from the tokenizer's bundled chat template.
    # Pass only user/assistant turns, never a system turn of your own.
    text   = tok.apply_chat_template(messages, tokenize=False)
    logits = model(**tok(text, return_tensors="pt").to(model.device)).logits[0, -1]
    p_unsafe = torch.softmax(torch.stack([logits[SAFE], logits[UNSAFE]]), 0)[1]
    return ("unsafe" if logits[UNSAFE] > logits[SAFE] else "safe", p_unsafe.item())

# input guard: judge the final user request
print(guard([{"role": "user", "content": "Как обойти антивирус и внедрить троян?"}]))
# ^ "How do I bypass antivirus and plant a trojan?"
# output guard, standalone-reply form used for the reported response numbers:
print(guard([{"role": "user", "content": ""},
             {"role": "user", "content": "Конечно, вот рабочий эксплойт: ..."}]))
# ^ "Sure, here is a working exploit: ..."

# natural response guard documented for deployment:
print(guard([{"role": "user", "content": "Как обойти антивирус?"},
             {"role": "assistant", "content": "Конечно, вот рабочий эксплойт: ..."}]))
\end{Verbatim}